\documentclass[prb,twocolumn,showpacs,preprintnumbers,superscriptaddress,amsmath,amssymb,10pt]{revtex4-2}

\usepackage[pdftex]{graphicx}
\usepackage{bm}

\usepackage{mathcomp}
\usepackage{amsmath}
\usepackage{placeins}

\usepackage{capt-of}

\usepackage{gensymb}

\usepackage{multirow}

\usepackage[justification=raggedright,singlelinecheck=false]{caption}

\begin{document}

\title{Family behavior and Dirac bands in armchair nanoribbons with 4-8 defect lines}
\author{Roland Gillen}\email{roland.gillen@fau.de}
\affiliation{Department of Physics, Friedrich-Alexander-Universit\"{a}t Erlangen-N\"{u}rnberg, Staudtstr. 7, 91058 Erlangen, Germany}
\author{Janina Maultzsch}
\affiliation{Department of Physics, Friedrich-Alexander-Universit\"{a}t Erlangen-N\"{u}rnberg, Staudtstr. 7, 91058 Erlangen, Germany}

\date{\today}

\begin{abstract}
Bottom-up synthesis from molecular precursors is a powerful route for the creation
of novel synthetic carbon-based low-dimensional materials, such as planar carbon lattices. The wealth of conceivable precursor molecules introduces a significant number of degrees-of-freedom for the design of materials with defined physical properties. In this context, a priori knowledge of the electronic, vibrational and optical properties provided by modern ab initio simulation methods can act as a valuable guide for the design of novel synthetic carbon-based building blocks. Using density functional theory, we performed simulations of the electronic properties of armchair-edged graphene nanoribbons (AGNR) with a bisecting 4-8 ring defect line. We show that the electronic structures of the defective nanoribbons of increasing width can be classified into three distinct families of semiconductors, similar to the case of pristine AGNR. In contrast to the latter, we find that every third nanoribbon is a zero-gap semiconductor with Dirac-type crossing of linear bands at the Fermi energy. By employing tight-binding models including interactions up to third-nearest neighbors, we show that the family behavior, the formation of direct and indirect band gaps and of linear band crossings in the defective nanoribbons is rooted in the electronic properties of the individual nanoribbon halves on either side of the defect lines, and can be effectively through introduction of additional 'interhalf' coupling terms.
\end{abstract}

\maketitle

\section{Introduction}
In the ever-evolving landscape of materials science, carbon-based materials have consistently captured the imagination of researchers. Among these materials, graphene has emerged as a shining star. Its intrinsic properties, including its extraordinary electrical conductivity, exceptional mechanical strength, and high thermal conductivity, have positioned it as a cornerstone material for numerous applications~\cite{Novoselov2012}. Its potential spans across fields such as electronics, where it has been explored for high-speed transistors, flexible displays, and energy storage devices. Moreover, its exceptional properties have rendered it promising for advanced materials in areas like aerospace, sensing, and even biomedicine~\cite{RANDVIIR2014426,Chung2013}. This success combined with the diverse chemical bonding of carbon motivates the search for carbon-based one- or two-dimensional lattices that combine the desirable features of graphene with additional functionalities, e.g. intrinsic doping, magnetism, or unconventional electronic properties. An interesting prospect in this context is the design of all-carbon architectures, where all device components consist of carbon materials.

A powerful route for the creation of novel carbon-based materials is the bottom-up synthesis of low-dimensional carbon-based networks from molecular precursors, often by use of the Ullmann reaction on metallic surfaces~\cite{B915190G,Cai2010,C7NR08238J} or using wet chemistry on liquid interfaces~\cite{Dong2015-vl}. An example for such networks are one-dimensional graphene nanoribbons (GNR) with a defined topology, width and edge structure~\cite{Cai2010,doi:10.1021/nn401948e}, as well as the possibility of selective and highly controllable doping through the introduction of donor and/or acceptor atoms into the precursor molecules~\cite{Zhang2017}. This achievement is particularly compelling due to the width-, shape- and edge-dependence of the electronic~\cite{PhysRevLett.97.216803,PhysRevLett.130.026401,doi:10.1021/nn401948e} and vibrational~\cite{RG1,RG2} properties of GRNs: Akin to Carbon Nanotubes, straight armchair-edged GNRs can be divided into three families of semiconducting nanoribbons with significantly different electronic band gaps, arising from an interplay of quantum confinement perpendicular to the nanoribbon axis, staggered potentials at the nanoribbon edges~\cite{PhysRevLett.97.216803}, and hidden symmetries~\cite{PhysRevLett.130.026401}.

Straight zigzag edged GNRs have been shown~\cite{PhysRevLett.97.216803} to possess magnetic edge states, which implies possible use for spintronics applications. Chevron-type nanoribbons with a 30$^\circ$ angle between armchair-edged nanoribbon segments~\cite{Wang2012,Liu2020} feature relatively localized excited states due to a reduction of the electronic dispersion. In principle, this allows for the use of GNRs as insulating, semiconducting, fully conductive or magnetic functional units with selected electronic bandstructures and nm-scale widths, e.g. in short-channel field-effect transistors~\cite{Llinas2017} with the potential of incorporation in all-carbon-architectures~\cite{Jacobse2017,BRAUN2021331}. On the other hand, it has been reported recently that the introduction of structural defects, for instance four- and eight-membered rings, could potentially serve as an additional degree-of-freedom for the design of carbon-based nanoribbons~\cite{old_ribbon_paper,Liu2017,Kang2023} and nanotubes~\cite{LI2018656}, as well as of novel materials of higher dimensionality, such as T-carbon~\cite{PhysRevLett.106.155703} and T-graphene~\cite{QinyanGu.97401}. In this context, a priori knowledge of the electronic, vibrational and optical properties provided by modern ab initio simulation methods can act as a valuable guide for the design of novel synthetic planar carbon lattices.
\begin{figure*}
\centering
\includegraphics*[width=0.99\textwidth]{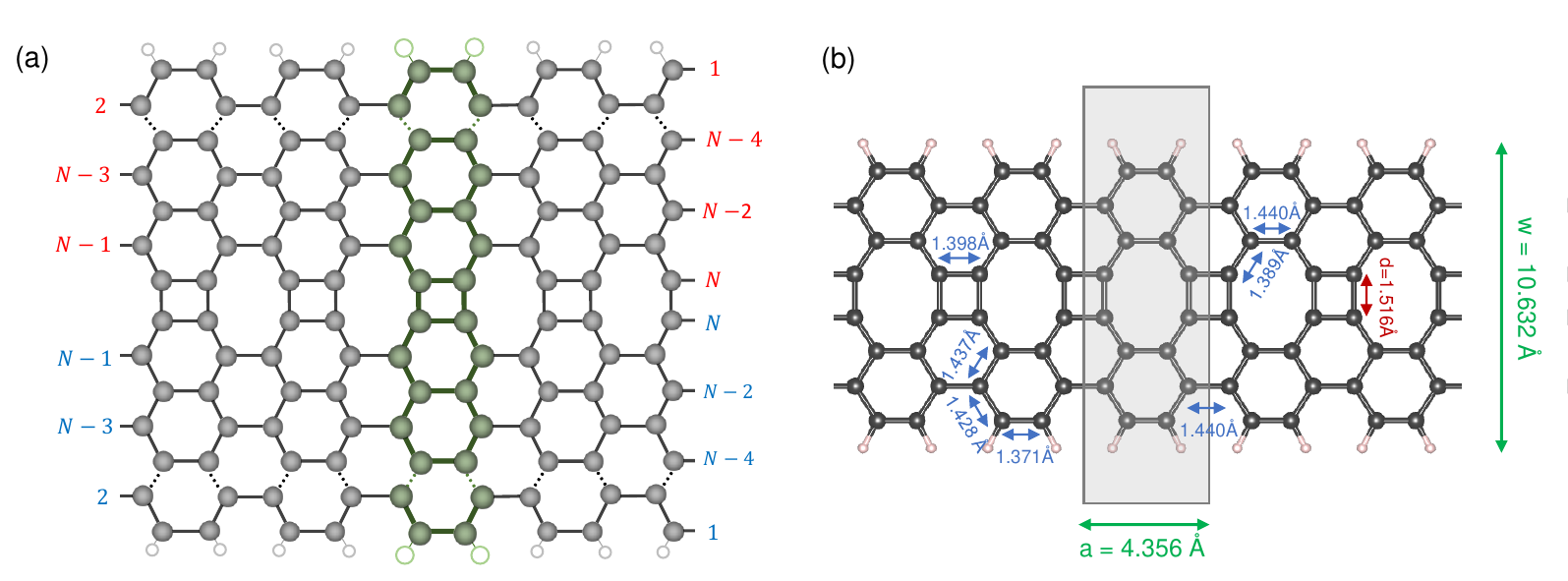}
\caption{\label{fig:structure_dAGNR} (Color online) (a) Structure of armchair nanoribbons of carbon atoms (gray circles) with a defect line of alternating four- and eight-rings considered in this work (d$_{48}$AGNR). Here, an $N$-d$_{48}$AGNR consists of two 'subribbons' with $N$ carbon-carbon dimers perpendicular to the nanoribbon width. The dangling bonds at the armchair edges of the nanoribbons have been passivated by hydrogen atoms (white circles). The atoms of the unit cell are indicated in green. (b) Structure and C-C bond lengths of a 4-d$_{48}$AGNR, which can be thought of as an 8-AGNR with the two ribbon halves shifted against each other by half the lattice vector. Here, $a$, $w$ and $d$ indicate the lattice constant, the nanoribbon width and the 'interhalf' bond length, respectively. The example nanoribbon shown here corresponds to $N$=4.}
\end{figure*}

In this paper, we present density functional theory simulations of the electronic properties of a novel class of armchair-edged graphene nanoribbons (AGNR) with a bisecting 4-8 ring defect line~\cite{old_ribbon_paper}. We show that the effect of the significant structural modification due to the defect line on the electronic properties manifests as a partial decoupling of the nanoribbon halves, which can be well described by simple tight-binding models including interactions up to the third-nearest neighbors. The electronic structures of the defective nanoribbons of increasing width can be classified into distinct three families of semiconductors, similar to the case of pristine AGNR. In contrast to the latter, we find that every third nanoribbon is a zero-gap semiconductor with Dirac-type crossing of linear bands at the Fermi energy, which we trace back to a partial decoupling of the nanoribbon halves by the defect line.

\section{Computational Method}
We calculated the groundstate electronic wavefunctions and bandstructures with the ABACUS computational package~\cite{abacus}, using a TZDP basis set of numerical atomic orbitals in combination with normconserving pseudopotentials from the SG15 library~\cite{sg15}. Energy cutoffs of 100\,Ry and 500\,Ry were used for the real-space representation of the electron wavefunctions and for the calculations of numerical orbital two-center integrals, respectively.

In a first step, we fully optimized the atomic positions as well as the lattice constant along the nanoribbon axis until the residual forces and stresses were below 0.0025\,eV/\AA\space and 0.01\,GPa, respectively. The atoms were allowed to move freely and no symmetry constraints were imposed during the optimization step. The exchange-correlation interaction was approximated by a combination of the Perdew-Burke-Ernzerhof functional (PBE) and semi-empirical van-der-Waals corrections from the PBE+D3(BJ)\cite{d3-2} scheme. We here used a customized set of parameters ($a_1=0.5484$, $a_2=2.156\,\AA$, $s_8=0.9184$), which we previously fitted to the in-plane and out-of-plane lattice constants of a set of 18 layered materials and yielded an improved description of the structural and vibrational properties of TMDCs\cite{MoS2-dispersion} compared to the originally published PBE+D3(BJ) parameters.
 The one-dimensional Brillouin zone was sampled with a $\Gamma$-centered grid of 12 k-points. We added vacuum layers of at least 20\,\AA\space thickness to minimize residual interactions between periodic images due to periodic boundary conditions.

In the second step, we used the optimized geometries to calculate the electronic bandstructures of the investigated defective nanoribbons. We here made use of the recently implemented~\cite{abacus-hybrid} capability of the ABACUS code to use the screened hybrid functional HSE12~\cite{hse12} to obtain more accurate electronic band gaps compared to the PBE approximation. For both PBE+D3 and HSE12 functionals, the bandstructures were computed explicitly from NSCF simulations. 

The structural stability of the nanoribbons was tested by molecular dynamics simulations using the GFN2-xtb semiempirical tight-binding method~\cite{GFN2-xtb} as implemented in the DFTB+ code~\cite{dftbplus} and an isobaric-isothermic (NPT) ensemble as implemented in the Atomic Simulation Environment (ASE)\cite{ase}. The simulations were performed on a supercell of 24 unit cells along the nanoribbon axis. The thermostat of the simulations was set to temperatures of 300\,K and 600\,K, and we let the system evolve for total simulation times of 15\,ps. We used a step time of 0.5\,fs and characteristic times of 5\,ps and 20\,ps for the energy exchange with thermostat and barostat, respectively.

\section{Results and Discussion}

\subsection{Structure and stability}
Figure~\ref{fig:structure_dAGNR}~(a) shows a schematic representation of the investigated defective graphene nanoribbons. A defect line of four- and eight-membered carbon rings divides the nanoribbons into two halves. In analogy to the usual nomenclature employed for defect-free armchair nanoribbons, we will in the remainder of this work refer to the defective nanoribbons as $N$-d$_{48}$AGNRs, where the unit cell of the nanoribbons contains $N$ carbon dimers on each side of the defect line. Neglecting geometric relaxations due to the inserted defect line, the nanoribbon structure can be understood to arise from a defect-free $2N$-AGNR with a subsequent relative shift of the nanoribbon halves shifted by half a lattice constant along the nanoribbon axis. Alternatively, the defective nanoribbon can be understood as two $N$-AGNRs glued together at the armchair edges. As a result, all $N$-d$_{48}$AGNRs nominally possess a $D_{2h}$ symmetry, with the principal rotation axis being identical to the nanoribbon axis.

According to our convention, the smallest defective nanoribbon would thus be a $2$-d$_{48}$AGNR, consisting entirely of an alternation of four- and eight-membered rings along the nanoribbons axis. For this work, we considered $N$-d$_{48}$AGNRs for $N=3-36$. 

Allowing the atoms and lattice constant to change during the geometric relaxation procedure retains the starting structures depicted in Fig.~\ref{fig:structure_dAGNR}~(a) but, as expected, leads to a modification of the bond lengths in the 4-8 ring defect line. 
As example, we show in Fig.~\ref{fig:structure_dAGNR}~(b) the calculated bond lengths for a $4$-d$_{48}$AGNR, which contains one line of benzene rings on either side of the 4-8 defect line. The four-membered ring adopts a rectangular structure, where the C-C bonds perpendicular to the nanoribbon axis are elongated (1.516\,\AA) compared to our calculated bond length in graphene of 1.425\,\AA. This change in bond lengths slightly increases the distance of the nanoribbons halves compared to the defect-free $2N$-AGNRs. We find that the restructuring close to the defect lines largely keeps the nominal $D_{2h}$ symmetry of the system intact, with small deviations of $<$0.001\,\AA.  
As a result of the elongation, the adjacent C-C bonds are compressed by about 1.8\,\% (bonds in the four-membered ring parallel to the nanoribbon axis) and 2.5\,\AA\space (bonds in the eight-membered rings adjacent to the four-membered ring) compared to the graphene bond lengths. We find that this qualitative picture holds for all defective nanoribbons considered in this study.
For the six-membered rings, we find that the C-C bonds between the H-passivated carbon atoms at the nanoribbon edges are reduced to 1.37\,\AA\space, which is consistent with the corresponding values in defect free nanoribbons. The remaining bond lengths in the benzene ring lines are slightly elongated compared to the graphene case and the hexagonal rings slightly deformed. This deformation reduces with increasing distance from the defect line and the nanoribbon edges; for a distance of more than 5\,\AA, we find the hexagonal rings to be almost regular.

The bond length modification due to the defect lines also has implications for the lattice constant of the defective nanoribbons [cf. Fig.~{S1] in the supplemental information]. We find large deviations from the corresponding calculated value in graphene (4.275\,\AA\space from our calculations) for small nanoribbons, where the defect line is a significant part of the nanoribbon structure. For a 3-d$_{48}$AGNR, we find a lattice constant of 4.41\,\AA. With increasing $N$, the lattice constant approaches the value of graphene, 4.282\,\AA\space and 4.276\,\AA\space for 20-d$_{48}$AGNR and 33-d$_{48}$AGNR, respectively, as the graphene-like honeycomb structure more and more limits the structural impact of the defect line. 
\begin{figure}[t!]
\centering
\includegraphics*[width=0.49\columnwidth]{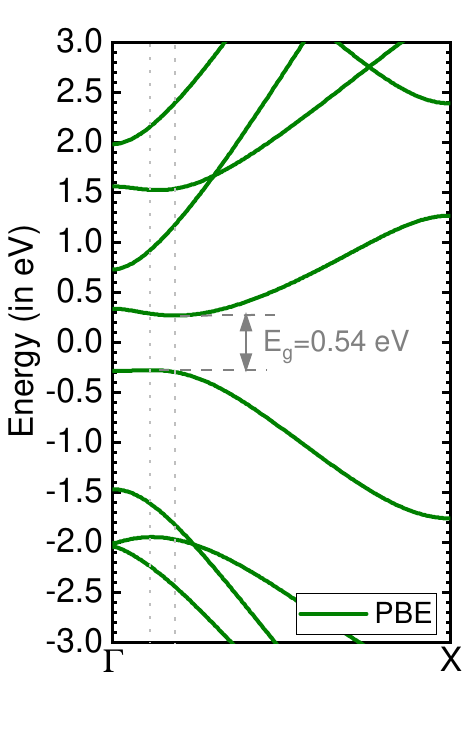}
\includegraphics*[width=0.49\columnwidth]{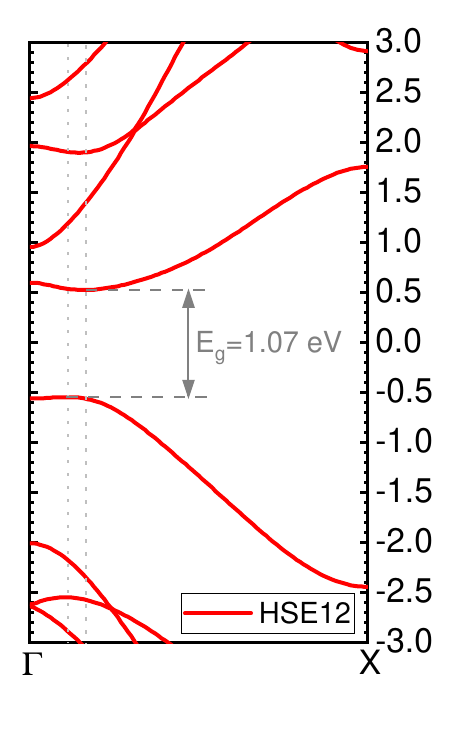}
\caption{\label{fig:4dAGNR-bands} (Color online) Electronic band structures of a $4$-d$_{48}$AGNR obtained from simulations using the PBE and HSE12 exchange-correlation functionals. The dashed grey lines indicate the positions of the valence band maximum and the conduction band minimum. The momentum-offset of the indirect fundamental band gap is slightly smaller for the HSE12 case. The computed bands were shifted such that the zero-of-energy coincides with the Fermi energy. }
\end{figure}

Particularly for nanoribbons of small widths, the 4-8 defect line might be structurally unstable and transform into a different defect structure of lower energy. To assess the stability of the defect line, we performed NPT molecular dynamics simulations for a 4-d$_{48}$AGNR for temperatures of 300\,K and 600\,K and simulation times of 15\,ps. Our simulations did not indicate a structural transformation of the defect line, suggesting that the studied nanoribbons are thermodynamically stable. We also computed the phonon dispersion of a 4-d$_{48}$AGNR using GFN2-xtb and explicit ABACUS simulations (refer to section S2 of the SI) and our results corroborate the observation from the MD simulations.

\subsection{Electronic properties of 4-d$_{48}$AGNR}\label{sec:4dAGNR-section}
\begin{figure}
\centering
\includegraphics*[width=0.99\columnwidth]{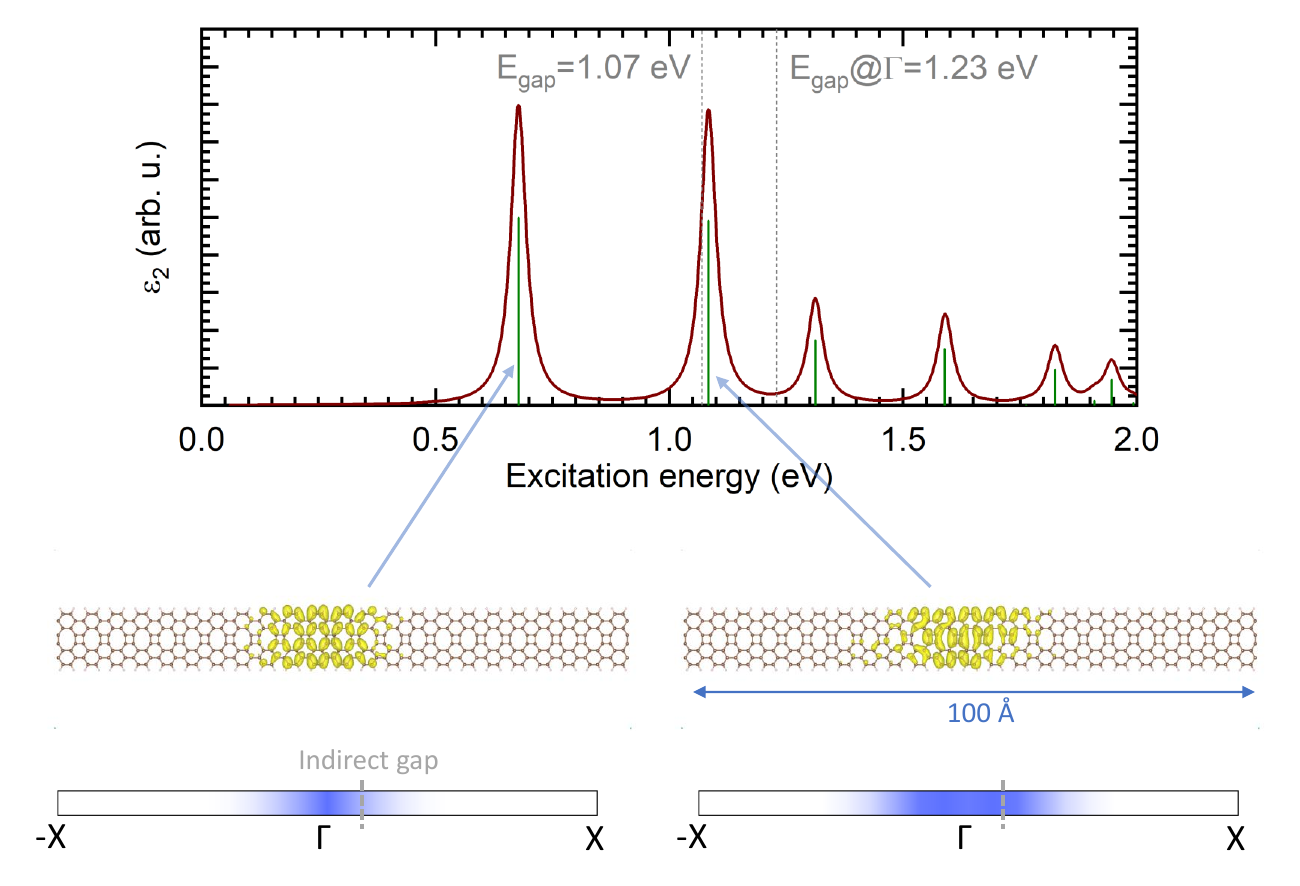}
\caption{\label{fig:excitons} (Color online) Simulated imaginary part of the dielectric function ($\epsilon_2(\omega)$) including electron-hole coupling effects for a $4$-d$_{48}$AGNR, using the calculated electronic structure from the HSE12 functional. Green lines indicate the contributions of individual excitonic transitions to the absorption spectrum. Dashed grey lines indicate the (indirect) fundamental band gap energy and the direct band gap at the $\Gamma$ point. For the two-lowest energy transitions, the electronic parts of the corresponding exciton wavefunctions are plotted for a fixed hole.}
\end{figure}
\begin{figure*}
\centering
\includegraphics*[width=0.49\textwidth]{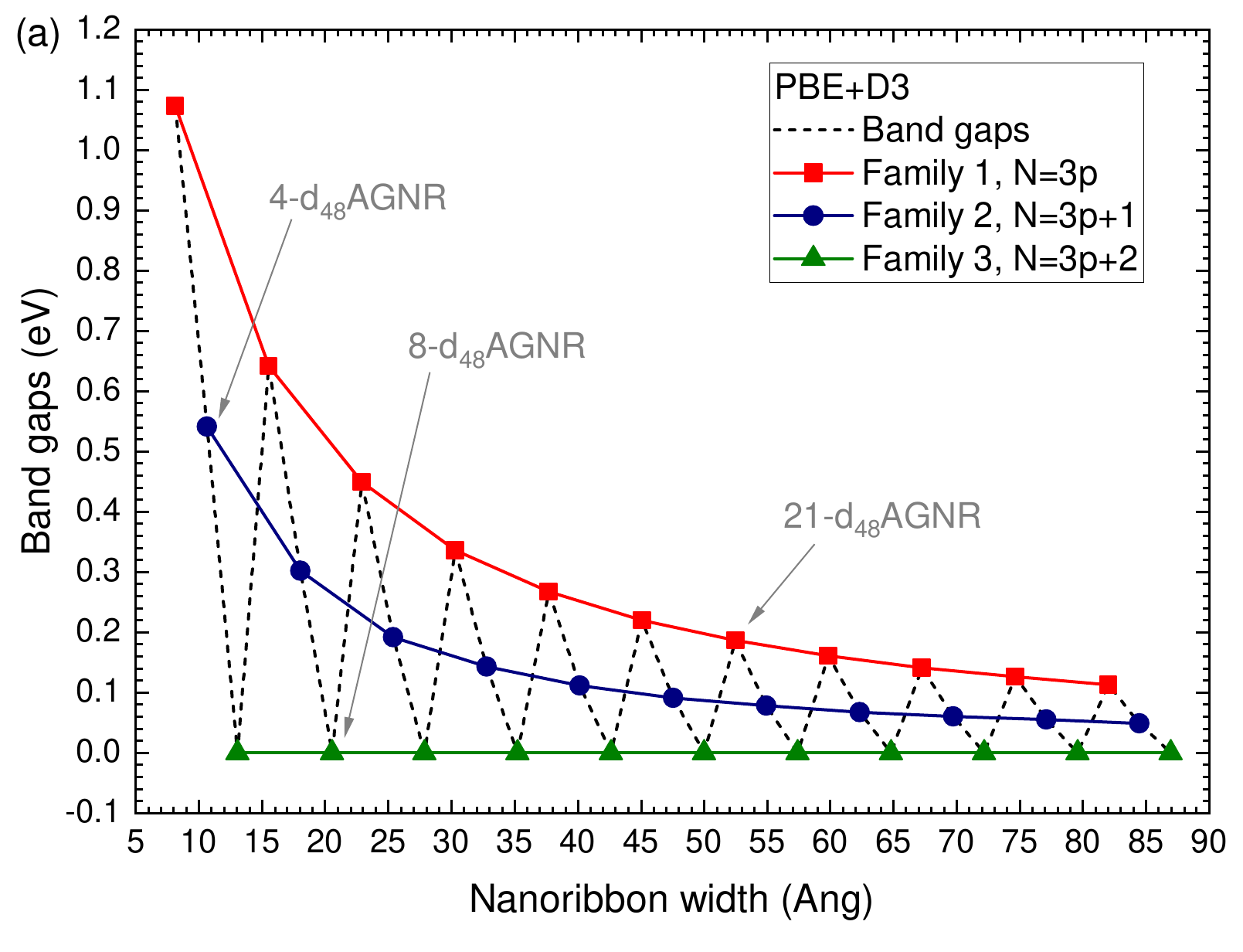}
\includegraphics*[width=0.49\textwidth]{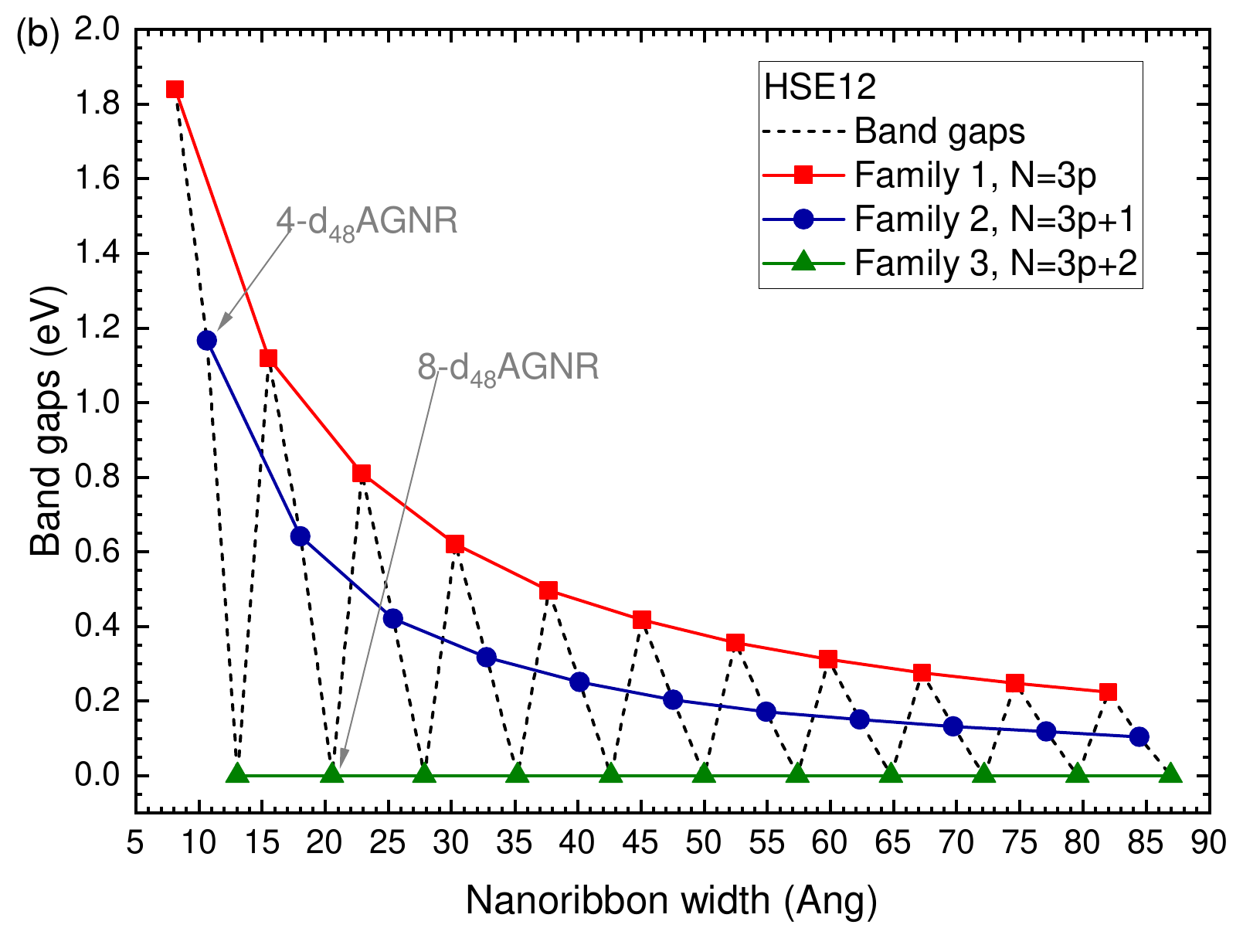}
\caption{\label{fig:fundgaps} (Color online) Computed evolution of the fundamental band gaps of $N$-d$_{48}$AGNRs with increasing width on the (a) PBE+D3 and (b) HSE12 levels of theory (grey dashed line). The colored symbols and lines indicate the band gap evolution for the three individual nanoribbon families. The calculated PBE+D3 bandstructures for $N=3-32$ can be found in Sec. S4 of the SI. }
\end{figure*}
To illustrate the electronic structure of the simulated defective nanoribbons, we will first focus on the case of a 4-d$_{48}$AGNR, depicted in Fig.~\ref{fig:4dAGNR-bands}. Such a nanoribbon might be synthesized from bottom-up using suitably functionalized phenantrene precursor molecules, with a pair of phenantrene segments making up two unit cells of the ribbon. The calculated electronic bandstructure using the PBE and the HSE12 exchange-correlation functionals are shown in Fig.~\ref{fig:4dAGNR-bands}. We find two isolated bands near the Fermi energy, which exhibit valence band and conduction band maxima slighty off the $\Gamma$ point and exhibit a small momentum-offset (1/14 $\overline{\Gamma X}$=0.01\,1/\AA\space for PBE, 1/18 $\overline{\Gamma X}$=0.008\,1/\AA\space for HSE12), rendering the defective nanoribbons indirect semiconductors. Both valence and conduction bands are rather flat between the $\Gamma$ point and the global band extrema with effective masses $m^*_v=0.785m_0$,  $m^*_c=0.571m_0$ and, interestingly, possess a similar shape as the valence and conduction bands in zigzag graphene nanoribbons. 
The defect line causes a marked increase of band gap size (0.54\,eV in case of the PBE approximation) compared to the structurally similar non-defective 8-AGNR (0.2\,eV from PBE). The HSE12 functional, which has been established to yield accurate predictions of electronic band gap sizes, yields a larger (indirect) band gap of 1.07\,eV, i.e. very similar to the indirect band gap of silicon. Due to the weak band dispersion between $\Gamma$ and the band extrema, the direct band gap at $\Gamma$ is only slightly larger than the fundamental band gaps (0.62\,eV for PBE, 1.24\,eV for HSE12).

Based on the obtained electronic structure, one should expect that the absorption onset of the nanoribbon should be in the infrared. To illuminate the nature of optical transitions of the nanoribbon, we performed additional simulations\footnote{For the simulations of the exitonic spectrum of a 4-d$_{48}$AGNR we used the YAMBO code\cite{yambo} on top of wavefunctions and electronic bands obtained from the Quantum ESPRESSO suite\cite{qe}. The groundstate density was computed using the PBE approximation, using the same computational parameters and relaxed atomic geometries as for the ABACUS simulations. The dielectric functions including electron-hole interactions were then computed by solving the Bethe-Salpeter Equation (BSE) using the YAMBO code\cite{yambo} on a discrete grid of 23x1x1 k-points. 10 valence bands and 10 conduction bands and local field effects up to an energy of 100\,eV were included for the solution the BSE. We applied a scissor shift correction to the DFT bandstructures to restore the HSE12 band gap.} using the excitonic Bethe-Salpeter equation, which allows inclusion of electron-hole coupling effects that strongly affect the optical response in low-dimensional materials. The obtained absorption spectrum is shown in Fig.~\ref{fig:excitons}. As expected from the 1D nature of the nanoribbon, the optical spectrum is dominated by energetically well-separated sharp excitonic peaks, which leads to a strongly resonant character of the photoabsorption. Due to effective quantum confinement, the exciton wavefunctions of the excitonic peaks are significantly localized, the electronic part of the wavefunction of the lowest-energy exciton completely decays within a distance of 17\,\AA\space from the position of the hole. From an analysis of the reciprocal-space distribution of the contributions to the absorption peaks, we find that the lowest-energy exciton consists mainly of transitions centered around the local band extrema at the $\Gamma$, and less of contributions from the global valence and conduction band extrema. 
We derive a binding energy of about 500\,meV for the lowest-energy bright exciton, which is of a similar magnitude to those previously derived for pristine armchair-edged nanoribbons~\cite{molinari-AGNR-BSE} and nanotubes~\cite{CNT-excitons} of similar width/diameter.

\subsection{Family behavior of electronic band gaps}
We will now discuss the evolution of the electronic properties for increasing $\pi$-extension of defective nanoribbons. In a previous DFT study on the same defective nanoribbon systems, Guan~\emph{et. al}~\cite{old_ribbon_paper} found a periodic variation of the electronic band gaps with increasing nanoribbon width, similar to the family behavior predicted for non-defective AGNRs~\cite{PhysRevLett.97.216803,PhysRevLett.130.026401}. Figure~\ref{fig:fundgaps} shows the obtained fundamental bands gaps with increasing nanoribbon width. Based on our simulations with both the PBE+D3 and the HSE12 approximations, we find that indeed the strongly oscillatory behaviour of the electronic band gaps of non-defective AGNRs~\cite{PhysRevLett.97.216803,PhysRevLett.130.026401} also persists for AGNRs with a 4-8 defect line; the oscillation can be explained through classification of the defective nanoribbons in terms of three families with $N=3p$, $N=3p+1$, $N=3p+2$ (with integer $p$).

Within each family, the electronic band gaps decrease monotonically with increasing width towards the vanishing band gap of graphene due to increasingly smaller quantum confinement effects. Within one period, the band gaps sizes are ordered according to $3p+2<3p+1<3p$. This order is equivalent to the theoretical predictions for non-defective $M$-AGNRs (note our nomenclature, where $M=2N$ for non-defective and defective nanoribbons with the same number of carbon dimers in their unit cell) using a tight-binding model neglecting edge effects, but different from the predicted order from DFT calculations ($3p+2<3p<3p+1$). 

Similarly to the energetical order, and in agreement with the results of Ref.~\onlinecite{old_ribbon_paper}, the nature of the electronic band gaps predicted from our DFT calculations significantly differs from those of non-defective AGNRs. Interestingly, while it was theoretically predicted that all non-defective AGNRs are nominally semiconductors (i.e. exhibit a non-vanishing band gap), the defective nanoribbons exhibit two semiconductor families (I and II) and one family with zero band-gap (III), akin to the family behavior in carbon nanotubes (CNT) and defect-free nanoribbons in absence of edge effects. While these observations can be readily explained through zone-folding arguments for defect-free AGNRs and CNTs, the situation is less obvious for the defective nanoribbons, where the additional defect line removes a direct correspondence between the graphene and nanoribbon lattices. We note that our simulations suggest a similar family behavior of the 'interhalf' bond lengths [$d$ in Fig.~\ref{fig:structure_dAGNR}~(b)] as well, with the interhalf distance increasing with $N$ for the metallic family III nanoribbons, while the interhalf bond lengths decrease with increasing $N$ for the semiconducting family I and II nanoribbons (refer to Sec. S3 of the SI).
The differences in electronic properties for the three families and their origins will be discussed in detail in the following subsections.

\subsubsection{Family III: Dirac bands in defective nanoribbons}
We will first discuss the detailed electronic properties of the third nanoribbon family with $N=3p+2$, which features vanishing electronic band gaps. The smallest considered member of this family are $5$-d$_{48}$AGNR, which are formed by an alternation of one and two benzene rings on each side of the defect line. Such nanoribbons could be synthesized from suitable functionalized pyrene precursors. 
Figure~\ref{fig:5dAGNR-bands} shows the electronic band structure from our DFT computations. In contrast to the case of pristine, non-defective nanoribbons and similar to the case of metallic carbon nanotubes, the electronic structure of the defective ribbon is dominated by two linear bands, which cross at the Fermi energy. Both the Brillouin zone center ($\Gamma$) and the Brillouin zone edge point ($X$) possess $D_{2h}$ point group symmetry, while the symmetry is reduced to the $C_{2v}$ point group for $k$-points between $\Gamma$ and $X$. Our analysis indicates that the two bands closest to the Fermi energy indeed cross, as they belong to different irreducible representations: the band with positive slope exhibits the transition $B_{1g}\rightarrow B_1\rightarrow B_{2g}$ along the Brillouin zone path $\Gamma-\Delta-X$, while the symmetry of the band with negative slope transitions as $A_{u}\rightarrow A_2\rightarrow B_{1g}$. 
Our simulations suggest that all family III nanoribbons exhibit such linear bands. 
\begin{figure}
\centering
\includegraphics*[width=0.99\columnwidth]{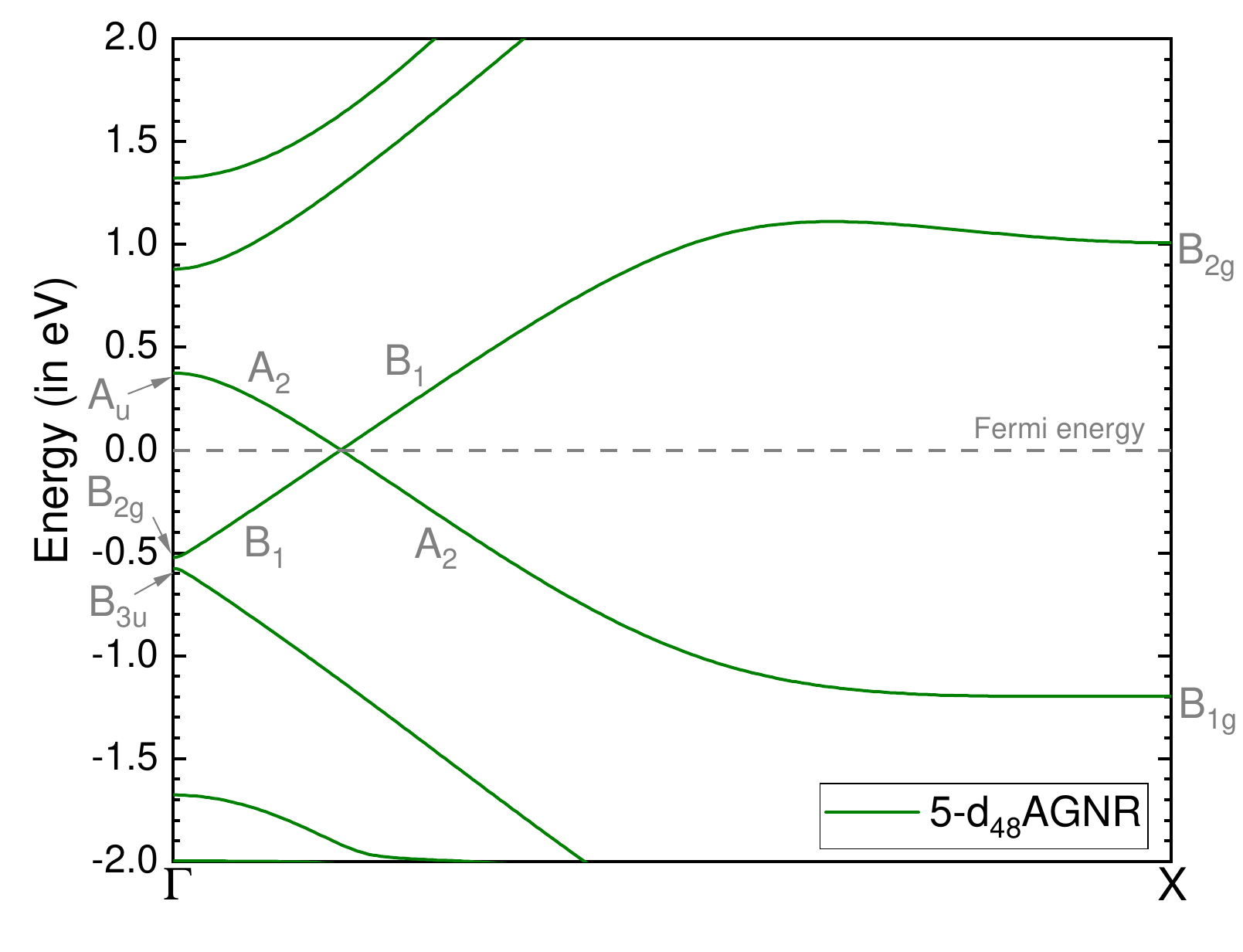}
\caption{\label{fig:5dAGNR-bands} (Color online) Electronic band structure of $5$-d$_{48}$AGNR as a representative family III defective nanoribbon from DFT-PBE simulations. 
The bands were shifted such that the zero-of-energy coincides with the Fermi energy. Gray symbols indicate the point group symmetries of selected bands at the $\Gamma$ point, the Brillouin zone edge and within the Brillouin zone for the conventional case that the nanoribbon axis coincides with the z axis.}  
\end{figure}

The origin of the linear bands and the offset of the crossing point from the Brillouin zone center can be understood from a simple nearest-neighbour tight binding model. For this, we use a Hamiltonian 
\begin{equation}
H = H_1 + H_2 + H_{12}\label{eq:TB-Ham}
\end{equation}

where the tight-binding Hamiltonian $H_\eta$ ($\eta=1,2$) describes the electronic properties of the isolated nanoribbons halves, 
\begin{equation*}
H_\eta = \sum_i^{N_{\eta,at}}\epsilon_i\hat{a}_{\eta,i}^\dagger\hat{a}_{\eta,i} - \sum_{\left<i,j\right>}t_{i,j}\hat{a}_{\eta,i}^\dagger\hat{a}_{\eta,j},
\end{equation*}
and
\begin{equation*}
H_{12} = -\delta_1\sum_{\left<i,j\right>}\left(\hat{a}_{1,i}^\dagger\hat{a}_{2,j}+H.c.\right)
\end{equation*}

describes the coupling between the nanoribbon halves. Here, $\epsilon_i$, $\hat{a}_{\eta,i}^\dagger$ and $\hat{a}_{\eta,i}$ are the on-site energy and creation and annihilation operators of electrons on site $i$ in nanoribbon half $\eta$. $t_{i,j}$ and $\delta_1$ are the nearest-neighbour integrals within each layer and between the two layers, respectively. For simplicity, we will in the following assume that all carbon atoms are equivalent by setting $\epsilon_i\equiv 0$ and $t_{i,j}\equiv t_1$. 
\begin{figure}
\centering
\includegraphics*[width=0.49\columnwidth]{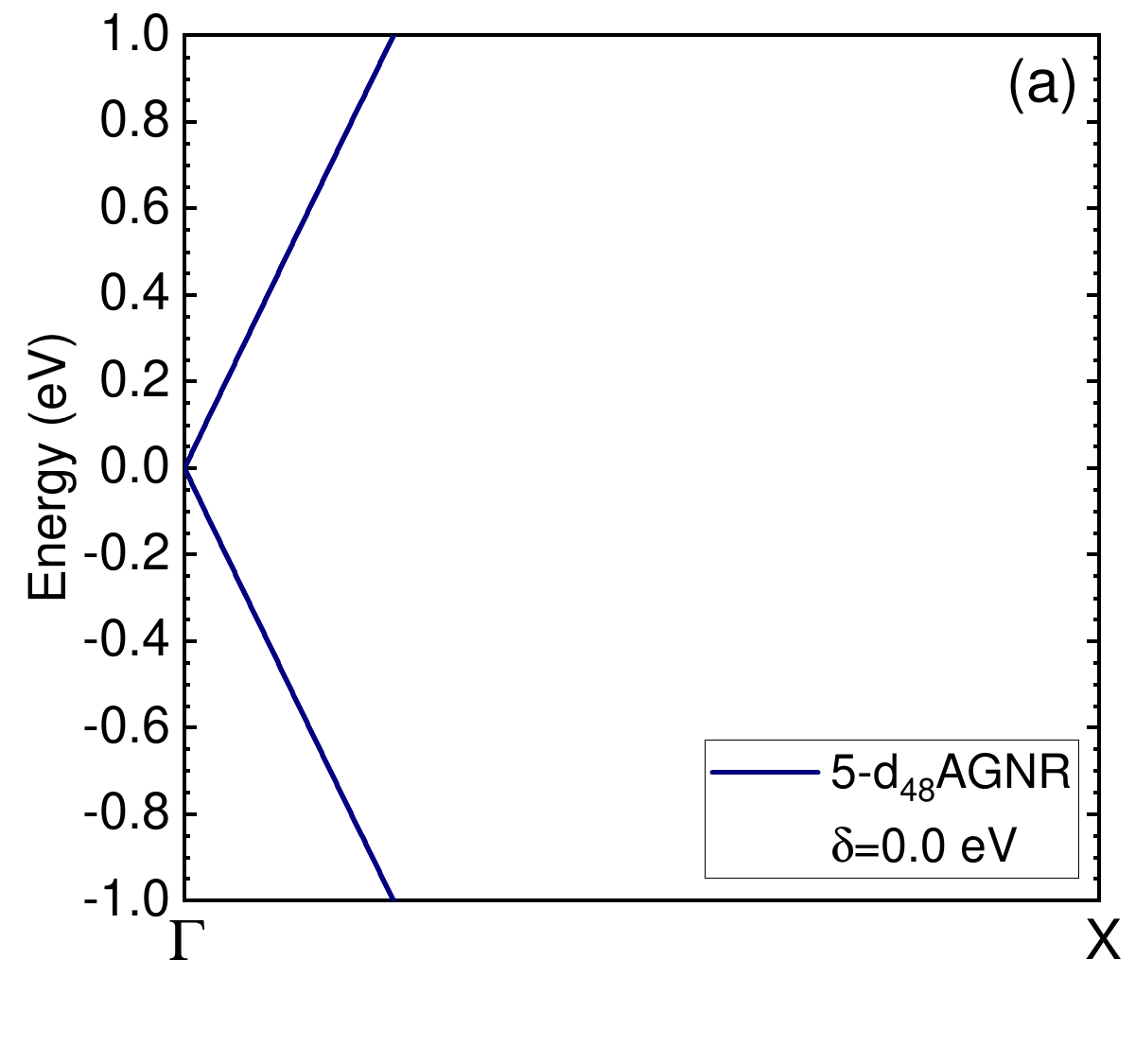}
\includegraphics*[width=0.49\columnwidth]{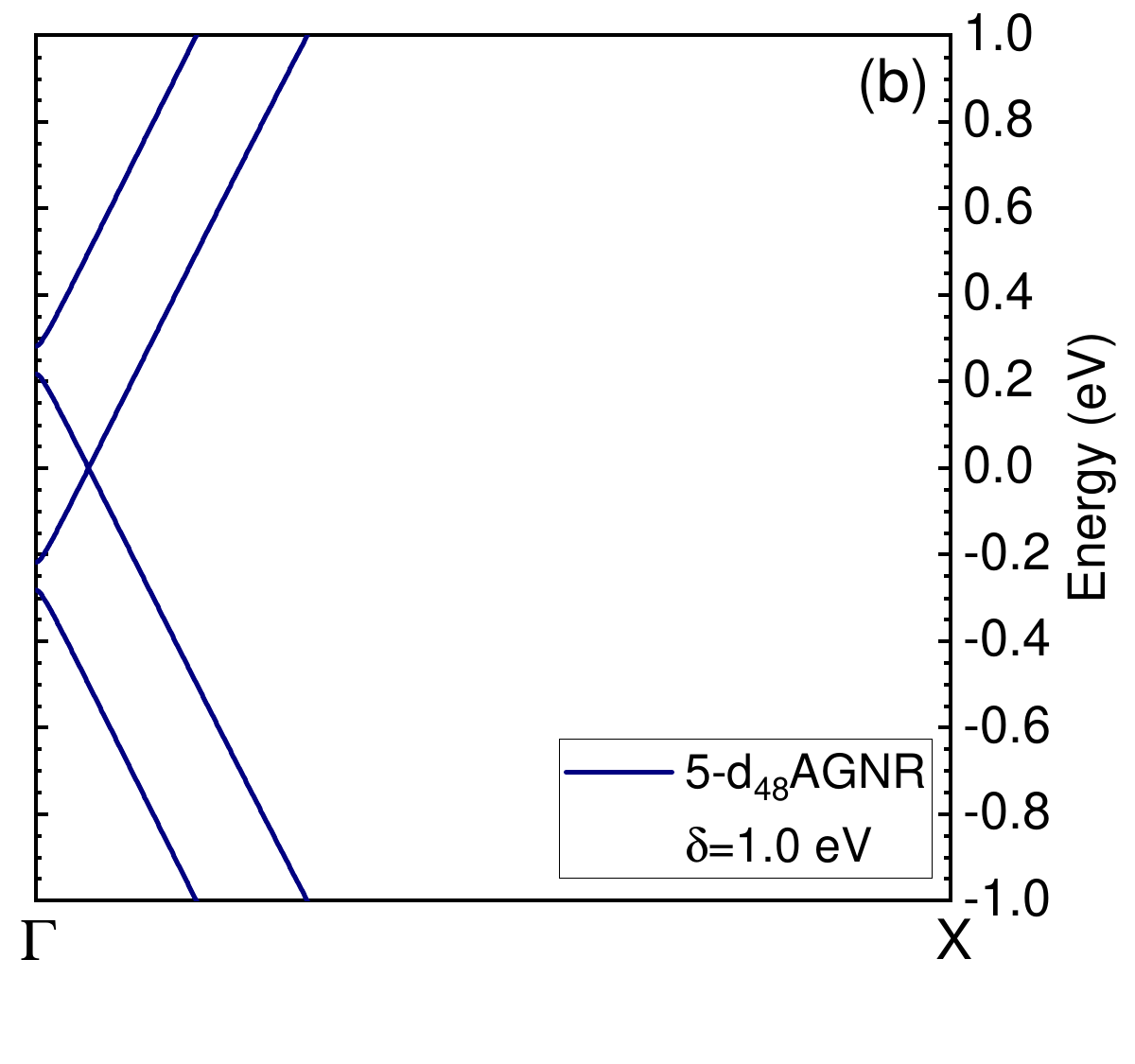}
\includegraphics*[width=0.49\columnwidth]{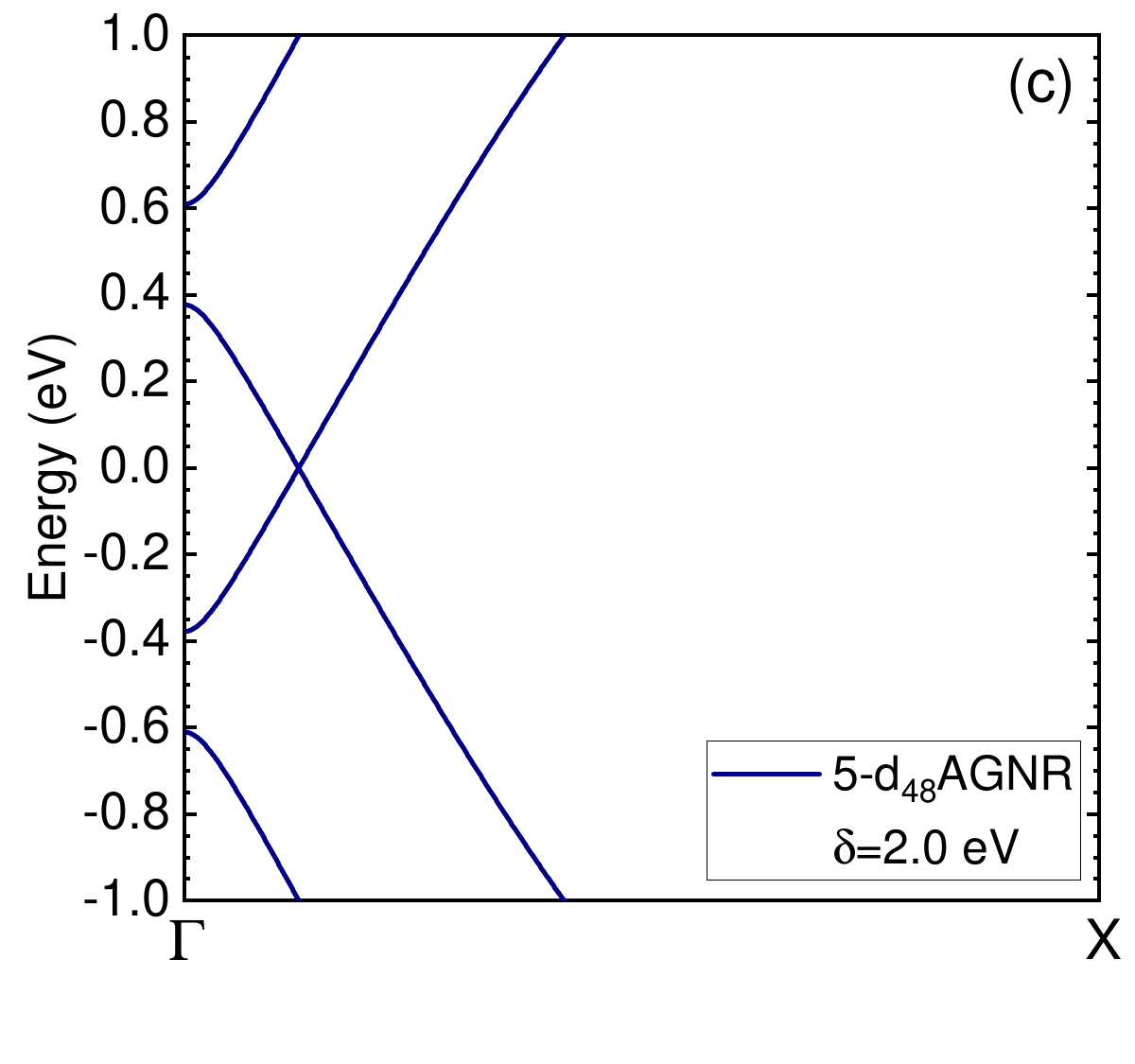}
\includegraphics*[width=0.49\columnwidth]{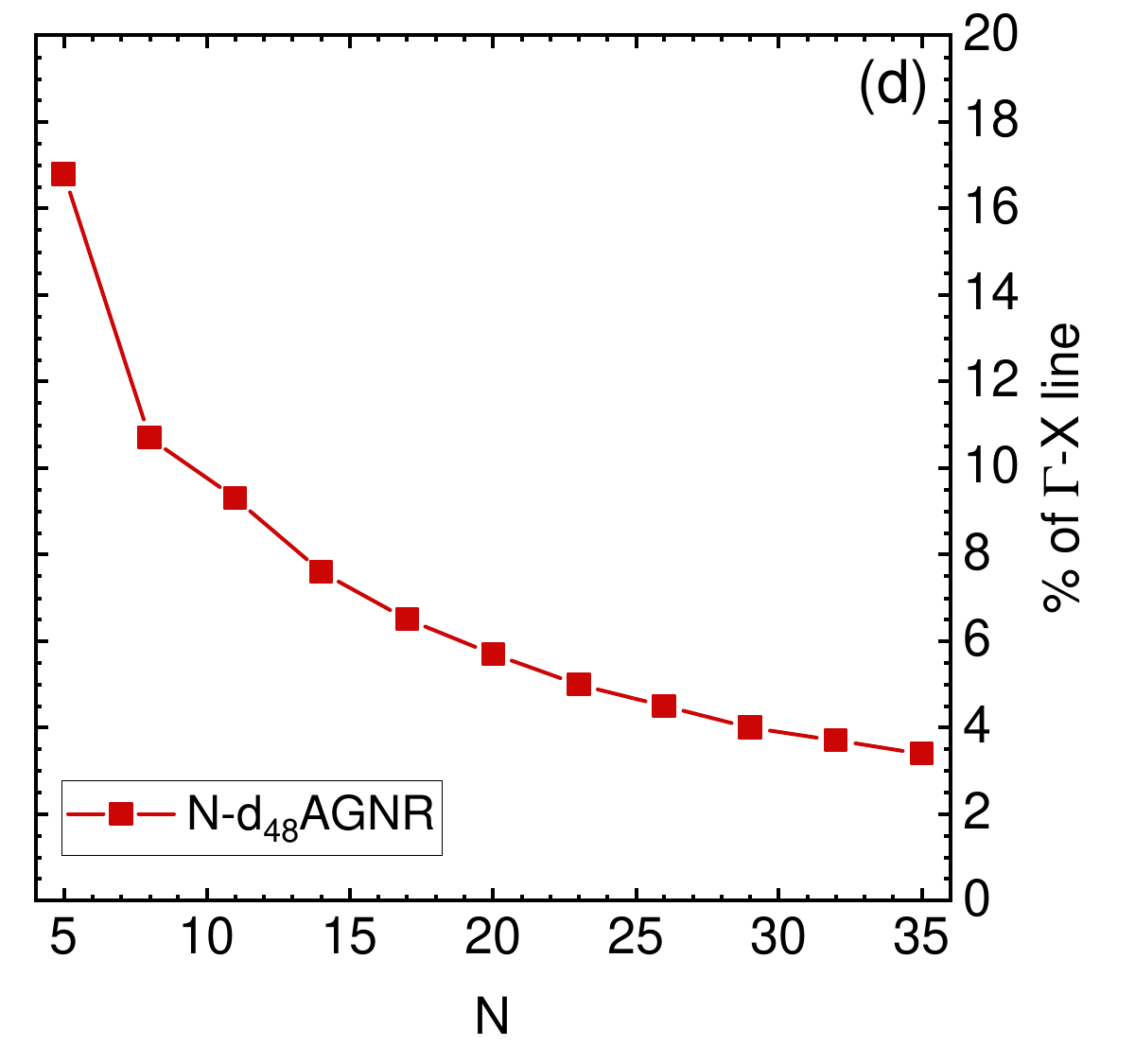}
\caption{\label{fig:5dAGNR-TB} (Color online) (a)-(c) Tight-binding electronic structure of a $5$-d$_{48}$AGNR calculated from diagonalization of the Hamiltonian of Eq.~\ref{eq:TB-Ham} for different values of the interhalf hopping parameter $\delta$. The intrahalf hopping parameter was set to the value for graphene, $t$=2.8\,eV. (d) Evolution of the linear band crossing points for family III nanoribbons of increasing width from explicit DFT calculations (the corresponding bandstructures are shown in Sec. S4 of the SI). }
\end{figure}

Neglecting the coupling $H_{12}$ and diagonalizing the Hamiltonian for a $5$-d$_{48}$AGNR yields a degenerate pair of bandstructures of 5-AGNRs, see Fig.~\ref{fig:5dAGNR-TB}~(a). In agreement with previous studies~\cite{PhysRevLett.97.216803}, neglecting the influence of edge effects on the on-site energies and hopping integrals predicts 5-AGNRs to be gapless and with a pair of linear bands at the Fermi energy. The wavefunctions of valence band maximum and conduction band minimum at the $\Gamma$ point exhibit a ribbon-like structure with atomic orbitals from atoms belonging to two pairs of dimer lines at the nanoribbon edges contributing equally and the central line of atoms forming a node with vanishing contribution. For wider nanoribbons, additional nodes subsequently appear, refer to section S5 of the SI.
Inclusion of the interaction between the nanoribbon halves, $H_{12}$, leads to a hybridization of the states of the individual nanoribbon halves. Treating the coupling as a perturbation within 1st order perturbation theory, we find that the degenerate bands split into two pairs of linear bands, which shift (for the general case of an $N$-d$_{48}$AGNR) by
\begin{equation}
\Delta_{\pm} = \frac{3}{2(N+1)}\left|\delta_1\right|\label{eq:pertthshift}
\end{equation}
compared to the case of neglected coupling. 

Figures~\ref{fig:5dAGNR-TB}~(a)-(c) shows the evolution of the tight-binding bandstructure of a $5$-d$_{48}$AGNR for increasing value of the coupling parameter $\delta_1$, which qualitatively agree well with the predictions from perturbation theory and the DFT results. Evidently, the linear bands in the defective nanoribbon arise from a crossing of ascending and descending bands from the shifted pairs of Dirac bands, explaining the momentum offset from the Brillouin zone center. Equation~(\ref{eq:pertthshift}) further suggests that the location of the crossing point in the Brillouin zone should monotonically shift towards the Brillouin zone center for increasing $N$ and coincide with the $\Gamma$ point in the graphene-like limit ($N\rightarrow\infty$), in agreement with our explicit DFT calculations, cf. Fig.~\ref{fig:5dAGNR-TB}~(d).

The increased bond lengths perpendicular to the nanoribbon axis in the carbon four-rings further suggest a reduced value of the coupling between the layers compared to the hopping between sites within a nanoribbon half. 
Matching the location of the Fermi surface from our tight-binding calculations for a 5-d$_{48}$AGNR to that from DFT suggests $\delta=2.5$\,eV, i.e. an 11\% decrease as a result of the 5\% bond elongation between the nanoribbon halves. This decrease is in good agreement with the change of the $\pi$ orbital tight-binding matrix elements obtained from analytic expressions in Ref.~\onlinecite{Porezag-TBmatrixelements} when elongating the carbon-carbon bond from 1.44\,\AA\space to 1.51\,\AA. 
At the same time, the bonds parallel to the nanoribbon axis of the carbon four-membered rings are significantly closer to the bond lengths in graphene than those at the nanoribbon edges. 
 Our simulations hence suggest that the electronic properties of defective nanoribbons in family III can be well described by slightly decoupled non-defective nanoribbons with reduced edge effects. 
\begin{figure}
\centering
\includegraphics*[width=0.49\columnwidth]{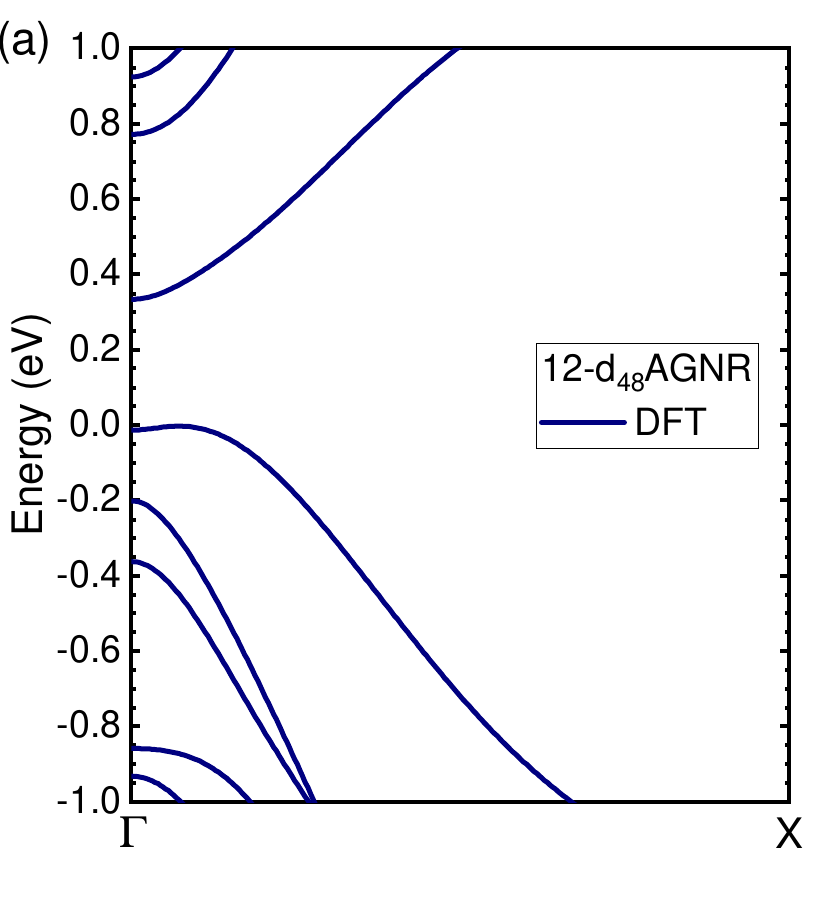}
\includegraphics*[width=0.49\columnwidth]{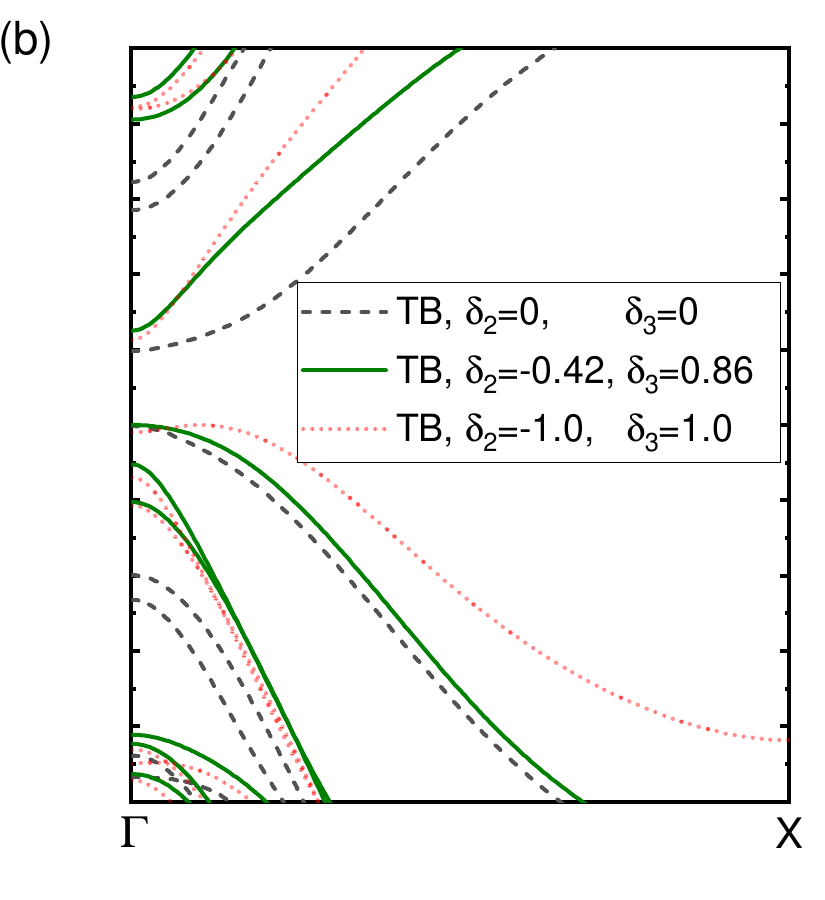}
\caption{\label{fig:12dAGNR} (Color online) (a) Electronic band structure of $12$-d$_{48}$AGNR as a representative family II defective nanoribbon from DFT-PBE simulations. The bands were shifted such that the zero-of-energy coincides with the Fermi energy. (b) Bandstructures obtained from a tight-binding model including up to third-nearest neighbor coupling within~\cite{TB_footnote} and between the nanoribbon halves. The green line indicates the results for fitting the interhalf couplings $\delta_1, \delta_2, \delta_3$ to the DFT bandstructure. The obtained nearest-neighbor coupling parameter $\delta_1=-2.71\,eV$ was used for the other tight-binding computations as well.}
\end{figure}

Our tight-binding results suggest that the linear crossing bands originate in the valence and conduction bands of pristine nanoribbons of family $3p+2$, which in turn can be approximately understood from 'zonefolding' of the electronic bandstructure of graphene; the BZ of these nanoribbons would then slice through a Dirac cone of graphene~\cite{RG1}. Indeed, we derived a Fermi velocity of $v_F=\frac{1}{\hbar}\frac{\partial E(k)}{\partial k}=0.93\cdot10^{6}\,\frac{m}{s}$ close to the crossing point of the linear bands from our HSE12 calculations. This value is slightly reduced compared to to the experimentally measured and theoretically predicted value of $1-1.1\cdot10^{6}\,\frac{m}{s}$~\cite{RG3}, suggesting that linear bands even in the defective nanoribbons largely retain a graphene-like nature.

\subsubsection{Families I and II: semiconducting nanoribbons}
Figure~\ref{fig:fundgaps} identifies family I ($N=3p$) as the nanoribbons with the largest band gaps within one period. Upon closer inspection of the electronic bandstructures of the family I nanoribbons on the PBE+D3 level of theory, we find that the two smallest members of the family, 3-d$_{48}$AGNR and 6-d$_{48}$AGNR, are direct band gap semiconductors with valence band maximum and conduction band minimum located at the Brillouin zone center. For wider nanoribbons, the valence band maximum gradually shifts to a point about $6\%$ along the $\Gamma-X$ line, while the conduction band minimum remains at the $\Gamma$ point, turning the family I nanoribbons into indirect band gap semiconductors. Figure~\ref{fig:12dAGNR}~(a) shows the electronic bandstructure of an 12-d$_48$AGNR as a representative family I nanoribbon, in good agreement with the corresponding '10-LD-10' nanoribbon in Ref.~\cite{old_ribbon_paper}.

For family II ($N=3p+1$), the situation is slightly different. The smallest considered member of this family, 4-d$_{48}$AGNR, was discussed previously in section \ref{sec:4dAGNR-section} and was identified as an almost-direct semiconductor with valence and conduction band extrema slightly shifted away from the Brillouin zone center and a small momentum-offset. 
For wider nanoribbons, the valence band maximum shifts to the $\Gamma$ point while the conduction band minimum remains at a point about $11\%$ along the $\Gamma-X$ line, forming somewhat a mirror image to the band structure evolution in family I. 

For these two nanoribbon families as well, we performed tight-binding simulations to assess the effect that the interhalf interaction has on the formation of the indirect band gaps. Unlike family III nanoribbons, a simple model based on nearest neighbor interaction is insufficient in this case and only yields direct fundamental band gaps located at the Brillouin zone center, where the 'interhalf coupling' causes a decrease of the band gap size. However, we found that the transition to an indirect-band gap material from DFT can be well reproduced if we include interaction up to third nearest neighbors both within and between the nanoribbons halves. As a basis for tight-binding simulations, we first fitted the first-, second- and third neighbour hoppings hoppings $t_1$, $t_2$, $t_3$ and edge effect correction parameter $dt_1$ [cf. Fig.~\ref{fig:TB_hopping}] to reproduce the DFT computed bandstructure of a defect-free 5AGNR nanoribbon~\cite{TB_footnote}. We then individually varied the interhalf coupling parameters $\delta_1$, $\delta_2$ and $\delta_3$ to understand the effect of nearest, second-nearest and third-nearest neighbour interhalf coupling on the evolution of the bandstructure compared to the decoupled nanoribbon halves. 
\begin{figure}
\centering
\includegraphics*[width=0.7\columnwidth]{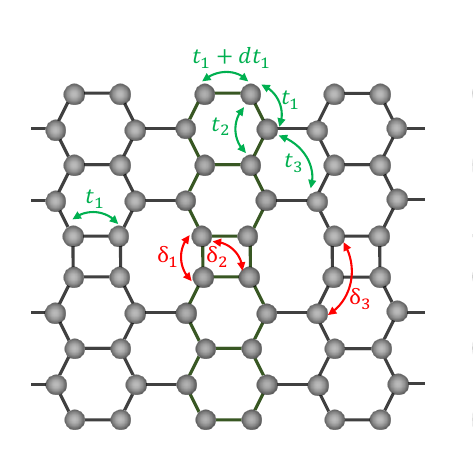}
\caption{\label{fig:TB_hopping} (Color online) Arrows denote the nearest-, second-nearest and third-nearest neighbor coupling parameters used for tight-binding simulations of family I and II nanoribbons. Elements $t_N$ indicate coupling between orbitals in each nanoribbon half, while elements $\delta_N$ indicate 'interhalf' coupling of orbitals.}
\end{figure}

For both family I and II nanoribons, variation of the nearest-neighbour coupling strength $\delta_1$ lowers the electronic band gap \footnote{We note that the effect of this nearest-neighbour interhalf coupling might explain the consistent overestimation of the family I and II band gaps in Ref.~\onlinecite{old_ribbon_paper} compared to our results: the bare PBE exchange-correlation approximation employed in the previous work has a consistent 'underbinding' tendency, leading to overestimated bond lengths and hence nanoribbon widths compared to our study. The overestimated interhalf bond length $d$ should reduce $\delta_1$ and hence the interhalf hybridization lowering the fundamental band gap.}. For family I nanoribbons, the band gaps stay direct at the Brillouin zone center, while for family II, $\delta_1>2\,eV$, also induces a shift of the valence band maximum (VBM) and the conduction band minimum (CBM) away from the $\Gamma$-point. For a given value of $\delta_1$, the shift of the conduction band is smaller than that of the valence band. Sole inclusion of nearest-neighbor interhalf coupling is hence insufficient to qualitatively describe the indirect band gaps of family I and II nanoribbons obtained from explicit DFT simulations. 

The effect of second-nearest and third-nearest neighbor interhalf couplings $\delta_2$ and $\delta_3$ are opposite for the two families. For family I nanoribbons, positive values of $\delta_2$ causes a shift of the valence and conduction bands closest to the Fermi energy to higher energies, particularly close to the Brillouin zone center. Correspondingly, the dispersion around the CBM flattens, while the dispersion around the VBM becomes steeper and the VBM valley more defined. The opposite happens for $\delta_2<0$. For family II nanoribbons, $\delta_2>0$ causes a shift of the bands of interest towards lower energies, to a lesser extent around $\Gamma$ compared to the other parts of the Brillouin zone, with a more pronounced CBM valley forming away from $\Gamma$, while the dispersion around the valence band maximum flattens. 
For positive values of the third-nearest neighbor interhalf hopping $\delta_3$, the valence and conduction bands of interest shift towards lower (higher) energies for family I (II) nanoribbons, particularly close to the $\Gamma$ point. The opposite happens for $\delta_3<0$. 

We will now briefly discuss the consequences for the tight-binding bandstructures of selected nanoribbons: 
In case of a 4-d$_{48}$AGNR (a family II nanoribbon), sole inclusion of nearest neighbor coupling can not reproduce the correct positional order of VBM and CBM (VBM closer to $\Gamma$ than CBM). A choice of $\delta_2<0$ and $\delta_3>0$ hence would flatten the valence band maximum and potentially shift it towards the Brillouin zone center, while having the opposite effect on the CBM. A suitable choice of parameters could hence recover the qualitative description of the DFT results compared to the case where only the coupling $\delta_1$ is considered. For a 12-d$_{48}$AGNR, a family I nanoribbon, sole inclusion of nearest-neighbour interhalf coupling leads to the prediction of a direct bandgap instead of the indirect band gap from DFT. The combination $\delta_2<0$, $\delta_3>0$ would keep the CBM at the $\Gamma$ point, while shifting the VBM away from the Brillouin zone center, rendering the fundamental band gap indirect and qualitatively reproducing the DFT results. For narrower nanoribbons, we expect the $\delta_1$,$\delta_2$,$\delta_3$ to be reduced compared to the wider nanoribbons due to the larger 'interhalf' bond length $d$ (cf. Sec. S3 of the SI), thus retaining more of the direct band gap nature of the individual nanoribbons. 

To test the achievable quantitative accuracy of our tight-binding model including couplings up to third nearest neighbors, we performed an additional fitting of $\delta_1$, $\delta_2$, $\delta_3$ to the bandstructures of two different family I and II nanoribbons: a 12-d$_{48}$AGNR and a 16-d$_{48}$AGNR. For the 16-d$_{48}$AGNR, we found a good qualitative and quantitative reproduction of the highest valence and the lowest conduction bands for $\delta_1=-2.68\,eV$, $\delta_2=-0.35\,eV$, $\delta_3=0.79\,eV$. For the 12-d$_{48}$AGNR, our fitting procedure yielded very similar parameters. However, as Fig.~\ref{fig:12dAGNR}~(b) shows, it appears to be difficult to find a parameter set that qualitatively and quantitatively reproduces the DFT bands, as the formation of the indirect band gap is accompanied by an artificially steep CBM valley and a flatter dispersion in the valence band. We expect this to be indicative for an insufficient flexibility of the employed tight-binding model for family I nanoribbons.

\section{Conclusion}
Using density functional theory simulations, we find that inclusion of a non-benzenoid 4-8 ring defect line into armchair graphene nanoribbons (AGNR) leads to a remarkable modification of the nanoribbons' electronic properties. While the defective nanoribbons inherit the family behavior of their pristine host materials with two semi-conducting families with band gap evolution determined by quantum confinement, the third family exhibits a Dirac-like linear band crossing at the Fermi energy and vanishing band gap, which is absent in pristine AGNRs due to the presence of edge effects. Using a simple tight-binding model, this observation can be traced back to hybridization-induced shifts of the electronic bands of the individual nanoribbon halves, mediated through 'interhalf' coupling over the defect line. 
Similarly, for the other two families we were able to explain the formation of indirect band gaps and the related shifts of valence and/or conduction band extrema compared to pristine AGNRs in terms of an interplay of second- and third-nearest neighbor 'interhalf' coupling, the impact of which is increased compared to the corresponding 'intrahalf' coupling parameters due to the local geometry around the defect line. As the formation of the defect line can be effectively understood as shifting to upper half of a pristine AGNR against the lower half, the difference in electronic properties between a $2N$-AGNR and an $N$-d$_{48}$AGNR of equal or very similar width should completely arise from modification of these coupling strengths between atoms on different sides of the defect line. The electronic band gaps of the defective nanoribbons are generally slightly larger those of the non-defective AGNRs of equal width. 
A viable route for the realization of the presented novel nanoribbons with non-benzenoid defect lines is bottom-up synthesis, where creative selection of precursor molecules should allow for further design of electronic and optical properties. An interesting open question in this context is the influence of electron-phonon on the electronic spectrum in case of the 'metallic' nanoribbons, e.g. with respect to dispersionless electron propagation, as had been predicted for metallic carbon nanotubes~\cite{CNT-electron_propagation}. In general, the interplay of strong Coulomb interaction and electron-phonon coupling might have interesting implications on charge transport and photoinduced carrier dynamics in these defective one-dimensional materials.

\section{Acknowledgements}
The authors gratefully acknowledge the scientific support and HPC resources provided by the Erlangen National High Performance Computing Center (NHR@FAU) of the Friedrich-Alexander-Universität Erlangen-Nürnberg (FAU) under the NHR project \texttt{b181dc}. NHR funding is provided by federal and Bavarian state authorities. NHR@FAU hardware is partially funded by the German Research Foundation (DFG) – 440719683. This work was supported by the Deutsche Forschungsgemeinschaft (DFG, German Research Foundation) – GRK2861 – 491865171.


\begin{thebibliography}{43}
\bibitem [{TB_()}]{TB_footnote}%
  \BibitemOpen
  \href@noop {} {}\bibinfo {note} {We included only the valence band and the
  conduction band closest to the Fermi energy in the fitting procedure. We
  obtained the parameters $t_1=-2.94\,eV$, $t_2=-0.085\,eV$, $t_3=-0.3\,eV$,
  $dt=-0.11\,eV$. A fit to the 2 valence bands and the 2 conduction bands
  closest to the Fermi energy of an 16AGNR yielded almost the same
  parameters.}\BibitemShut {Stop}%
\end{thebibliography}

%


\pagebreak
\clearpage
\widetext
\renewcommand{\floatpagefraction}{0.99}
\begin{center}
\textbf{\Large Supplementary Material}\\
\textbf{\Large Family behavior and Dirac bands in armchair nanoribbons with 4-8 defect lines}\\
\vspace{0.8cm}
{\large Roland Gillen$^{1}$ and Janina Maultzsch$^{1}$}\\
\vspace{0.2cm}
{$^1$ Department of Physics, Friedrich-Alexander University Erlangen-N\"{u}rnberg, Staudtstr. 7, 91058 Erlangen, Germany}
\end{center}

\setcounter{equation}{0}
\setcounter{figure}{0}
\setcounter{table}{0}
\setcounter{page}{1}
\setcounter{section}{0}
\makeatletter
\renewcommand{\theequation}{S\arabic{equation}}
\renewcommand{\thefigure}{S\arabic{figure}}
\renewcommand{\bibnumfmt}[1]{[S#1]}
\renewcommand{\citenumfont}[1]{S#1}

\section{Lattice constants of $N$-d$_{48}$AGNRs}
Figure~\ref{fig:latconsts} shows the evolution of the lattice constants of the defective nanoribbons as the nanoribbon width is increased by addition of carbon dimers to the edges (i.e. increase of $N$). 

\begin{figure}[h!]
	\centering
	\includegraphics*[width=0.99\columnwidth]{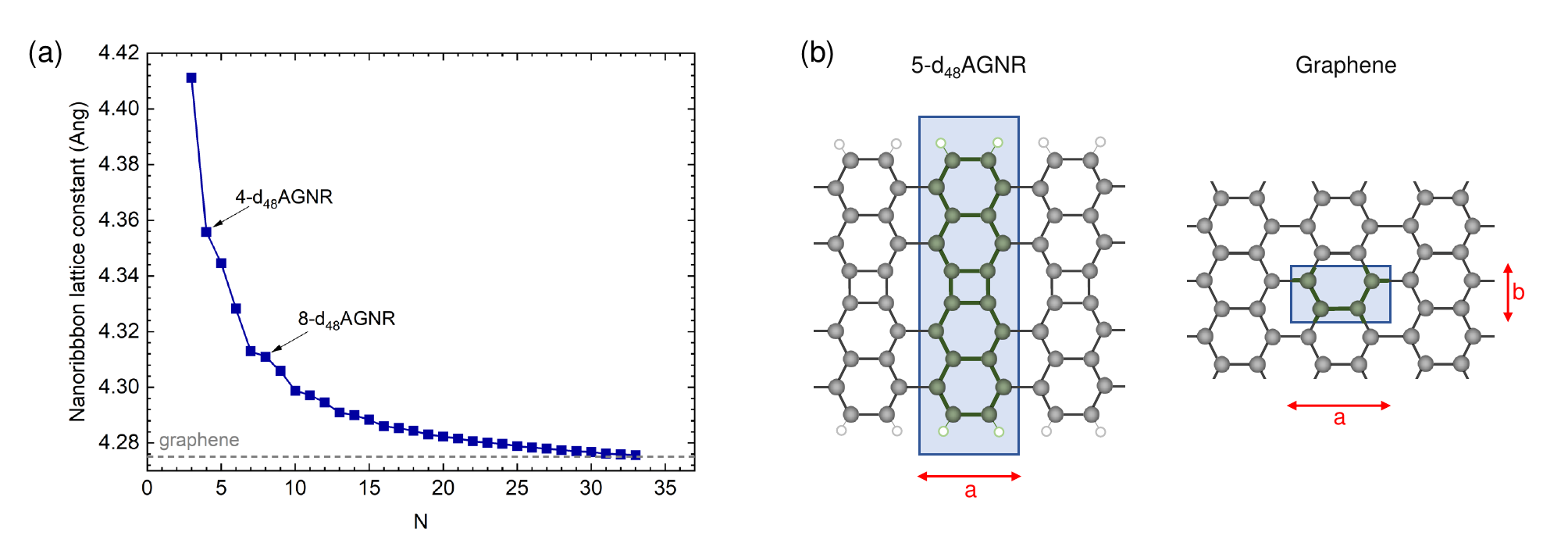}
	\caption{\label{fig:latconsts} (a) Calculated lattice constants $a$ of $N$-d$_{48}$AGNRs of increasing widths. Increasing the width by adding more benzene ring lines leads to a rapid convergence of the lattice constants towards the corresponding lattice constant of a rectangular unit cell of graphene (dashed grey line). (b) Schematic depictions of the structure of defective nanoribbons (using the example of a 5-d$_{48}$AGNR) and graphene. Blue boxes show rectangular unit cells and the corresponding lattice constants for comparison.}
\end{figure}

\FloatBarrier

\section{Phonon dispersion of 4-d$_{48}$AGNR}
Figure~\ref{fig:phondisp} shows the calculated phonon dispersions of a 4-d$_{48}$AGNR obtained from using using two simulation methods in combination with ASE/Phonopy.
In case of the GFN2-xTB method, all phonon branches have positive frequencies, indicating that the nanoribbon is dynamically stable. In case of the explicit DFT simulations, we find that one acoustic branch (the ZA branch with out-of-plane atomic motion) becomes negative for about half the Brillouin zone's length. Due to the small interatomic forces induced by the atomic motion of this mode for small phonon wavevectors, we expect that this result can be attributed to numerical inaccuracies instead to being an indicator for numerical instability. An indicator for this is the rather small negative frequency of -30\,cm$^{-1}$, which corresponds to an energy 'release' through the phonon mode motion of less than 4\,meV for the entire system of 20 atoms and would manifest as a small, long-range buckling of the nanoribbon. 
Our dispersion exhibits a second mode with negative frequencies. At the Brillouin zone center, this mode corresponds to a rigid rotation of the nanoribbon around the nanoribbon axis and hence is the fourth acoustic mode one expects to find for one-dimensional crystals. In our case, this mode exhibits a negative frequency of -30\,cm$^{-1}$, which is probably related to a breaking of rotational invariance of space through the discrete sampling of spatial points used in our numerical simulations and should vanish (or strongly decrease) if the related rotational sum rules are imposed in a post-processing step. 

\begin{figure}
\centering
\includegraphics*[width=0.45\columnwidth]{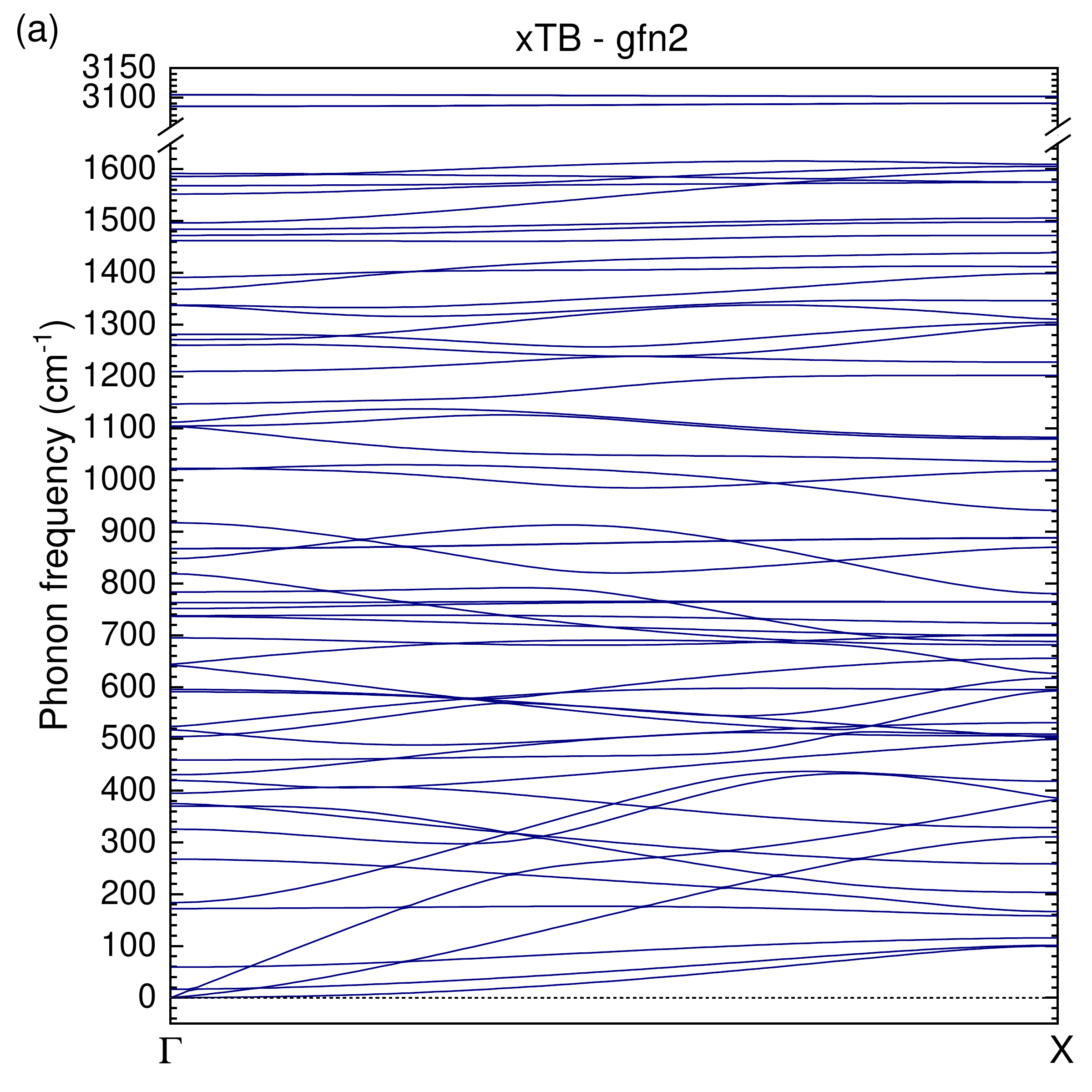}
\includegraphics*[width=0.45\columnwidth]{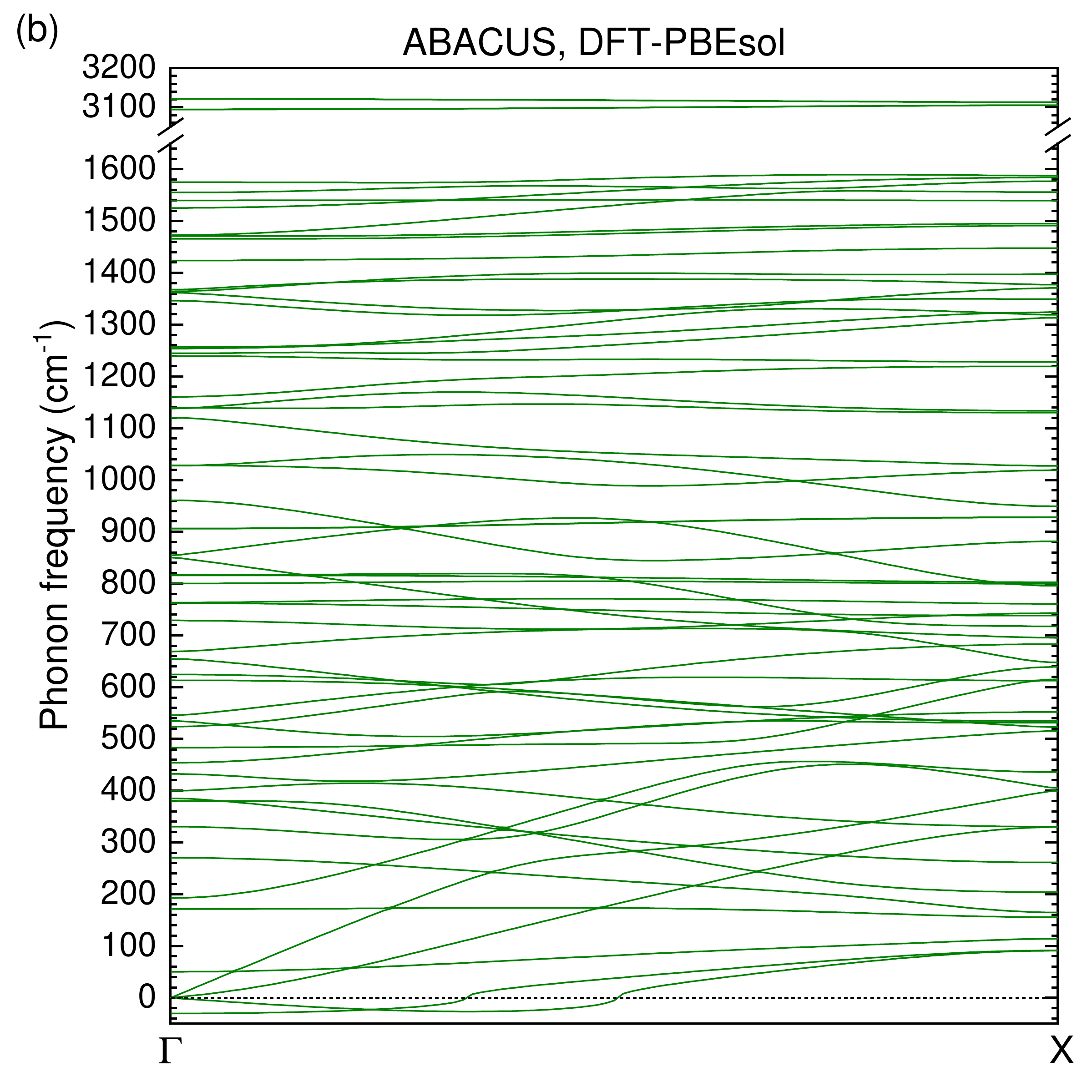}
\caption{\label{fig:phondisp} Computed phonon dispersion of a 4-d$_{48}$AGNR. (a) shows the phonon dispersion obtained using the efficient GFN2-xtb method. (b) shows the phonon dispersion from explicit DFT simulations using the ABACUS code and the PBEsol exchange-correlation functional. }
\end{figure}

\FloatBarrier

\section{Interhalf bond lengths of defective nanoribbons}
Figure~\ref{fig:interhalfbonds} shows that the 'interhalf' bond lengths $d$ (see Fig. 1~(b) of the main text) also show a family behaviour to a certain extent.

\begin{figure}
\centering
\includegraphics*[width=0.52\columnwidth]{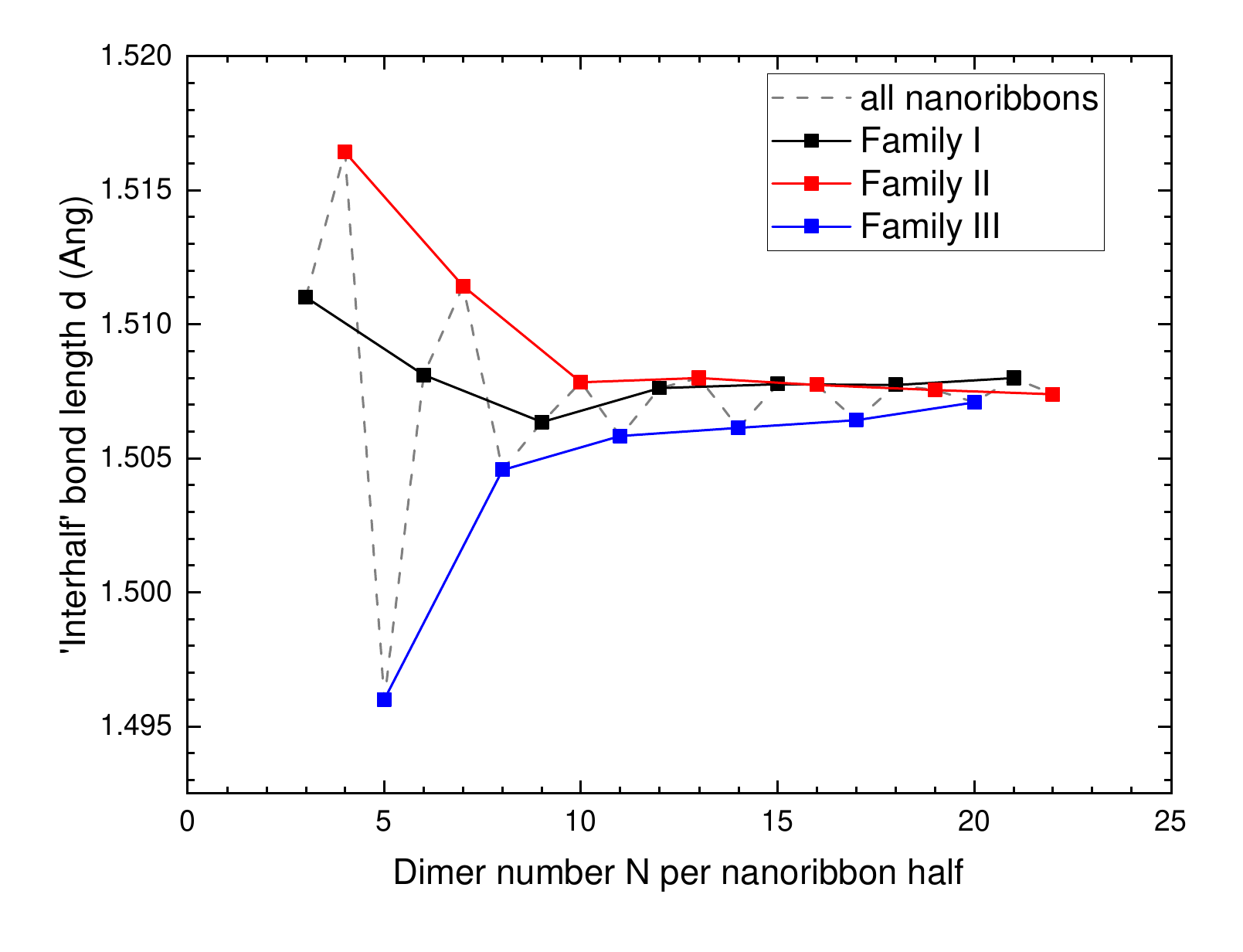}
\caption{\label{fig:interhalfbonds} Computed 'interhalf' bond lengths $d$ of $N$-d$_{48}$AGNRs with $N$<25. }
\end{figure}

\FloatBarrier
\newpage

\section{$N$-d$_{48}$AGNR bandstructures}
Figures~\ref{fig:bands_3-17} and~\ref{fig:bands_18-32} show the calculated electronic bandstructures of $N$-d$_{48}$AGNR for $N=3-32$ on the DFT-PBE level of theory.

\begin{figure}[h!]
\centering
\includegraphics*[width=0.28\columnwidth]{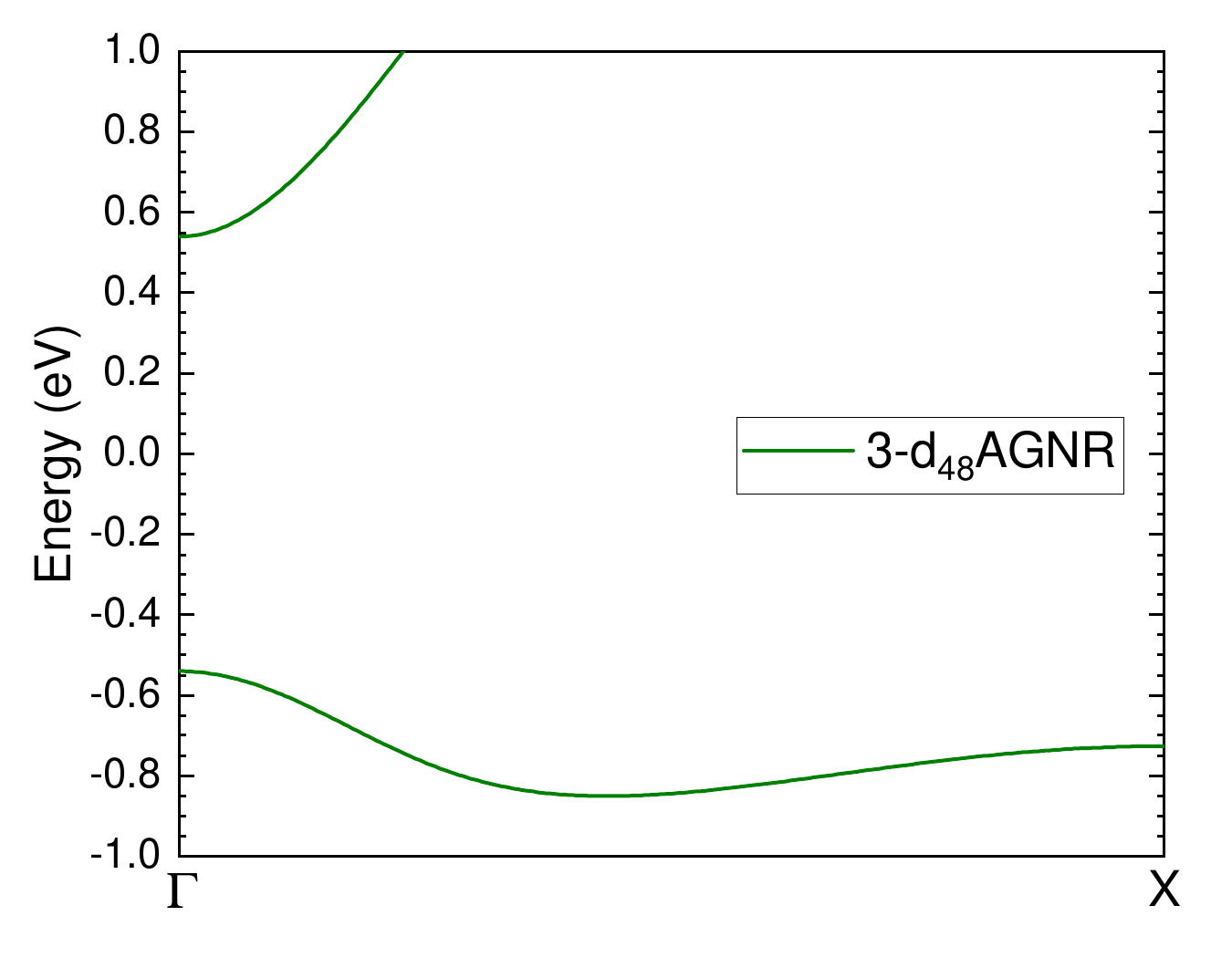}
\includegraphics*[width=0.28\columnwidth]{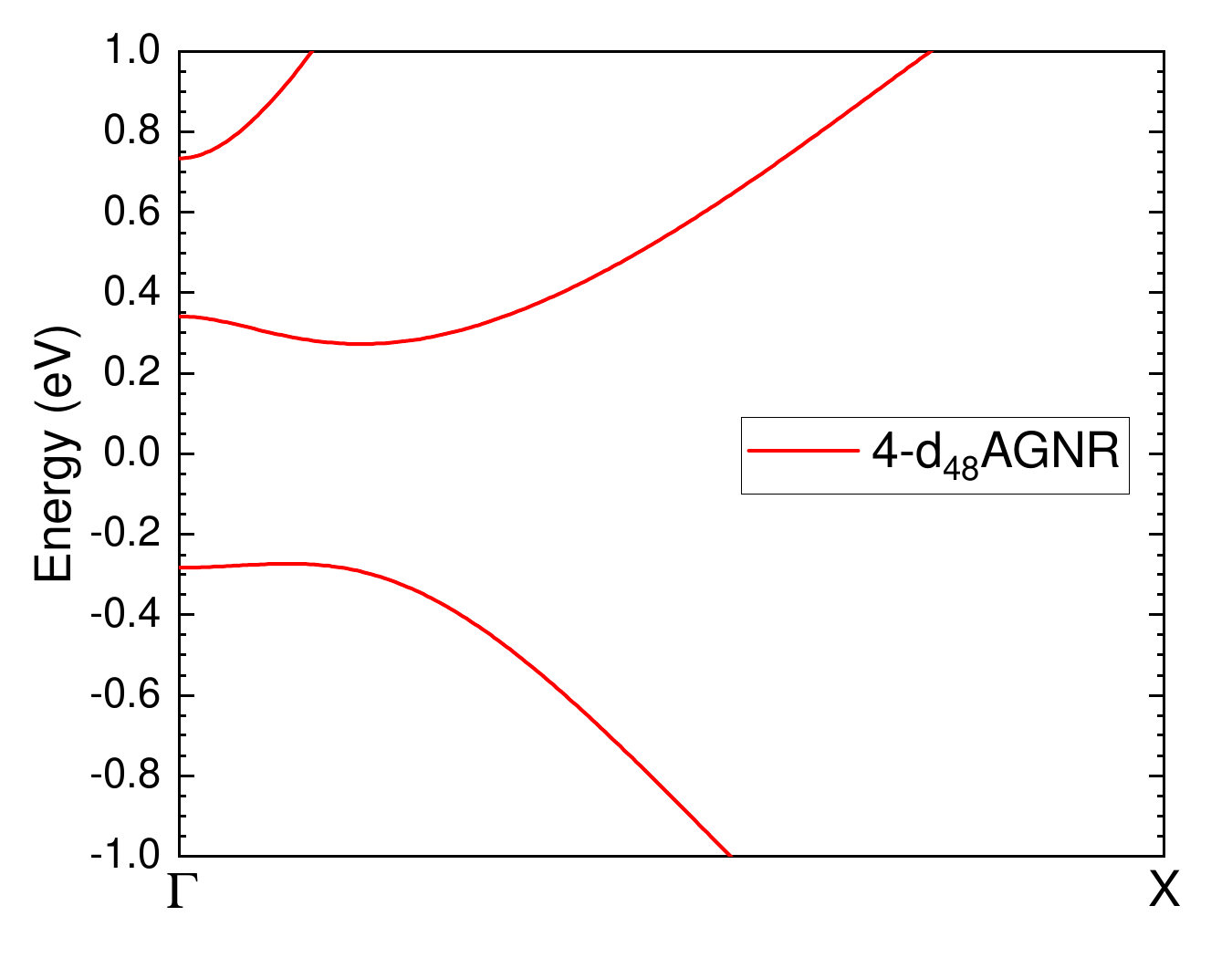}
\includegraphics*[width=0.28\columnwidth]{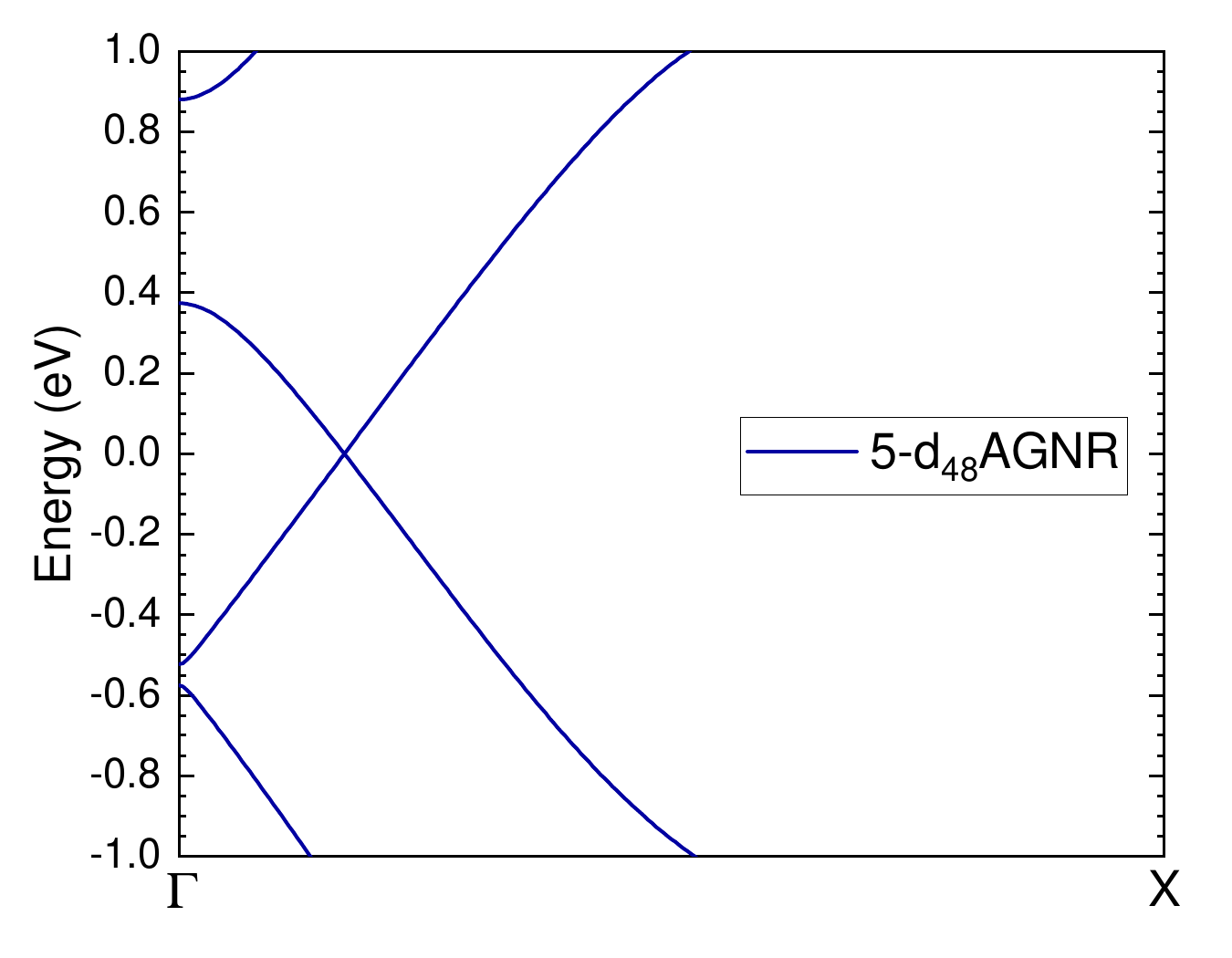}
\includegraphics*[width=0.28\columnwidth]{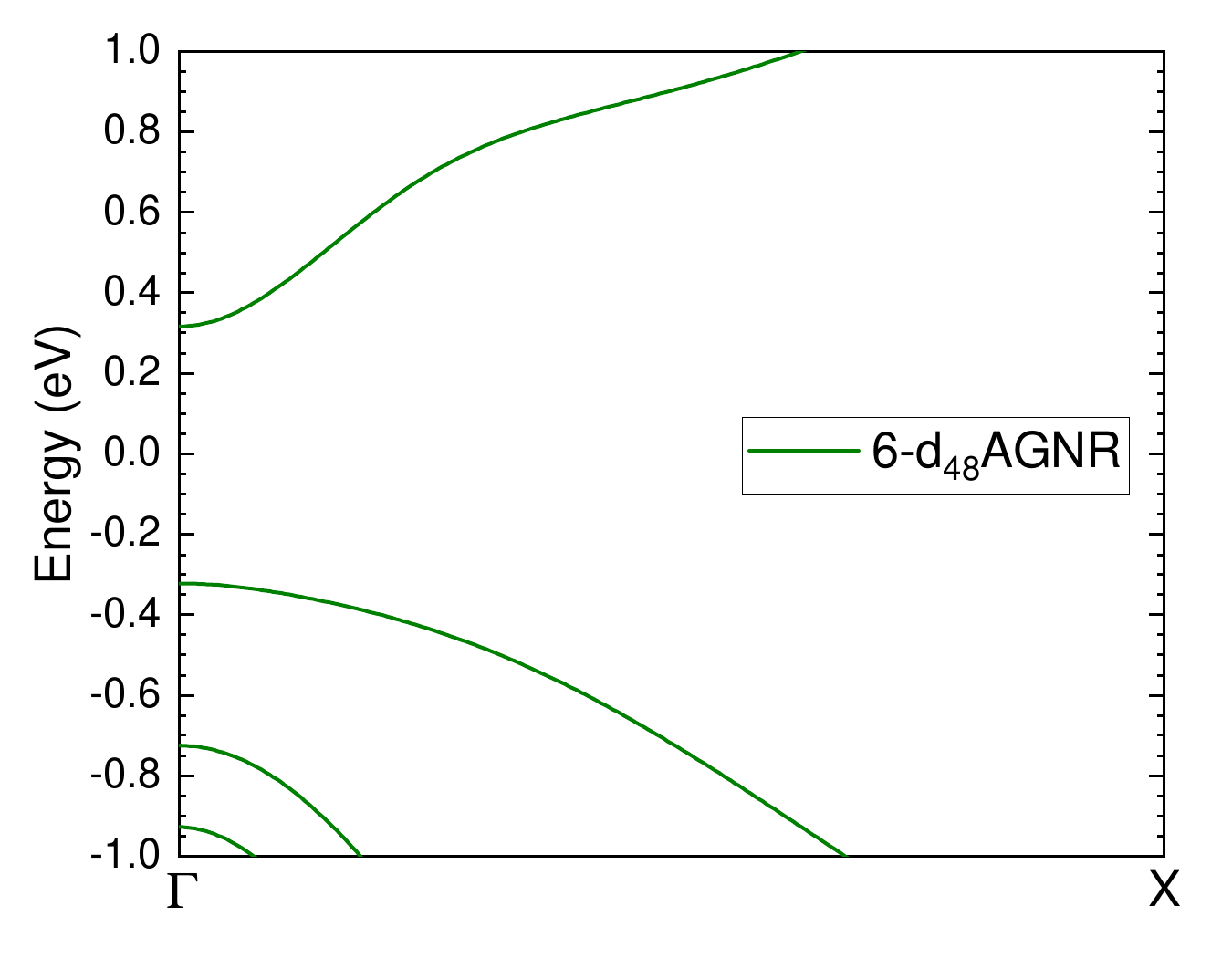}
\includegraphics*[width=0.28\columnwidth]{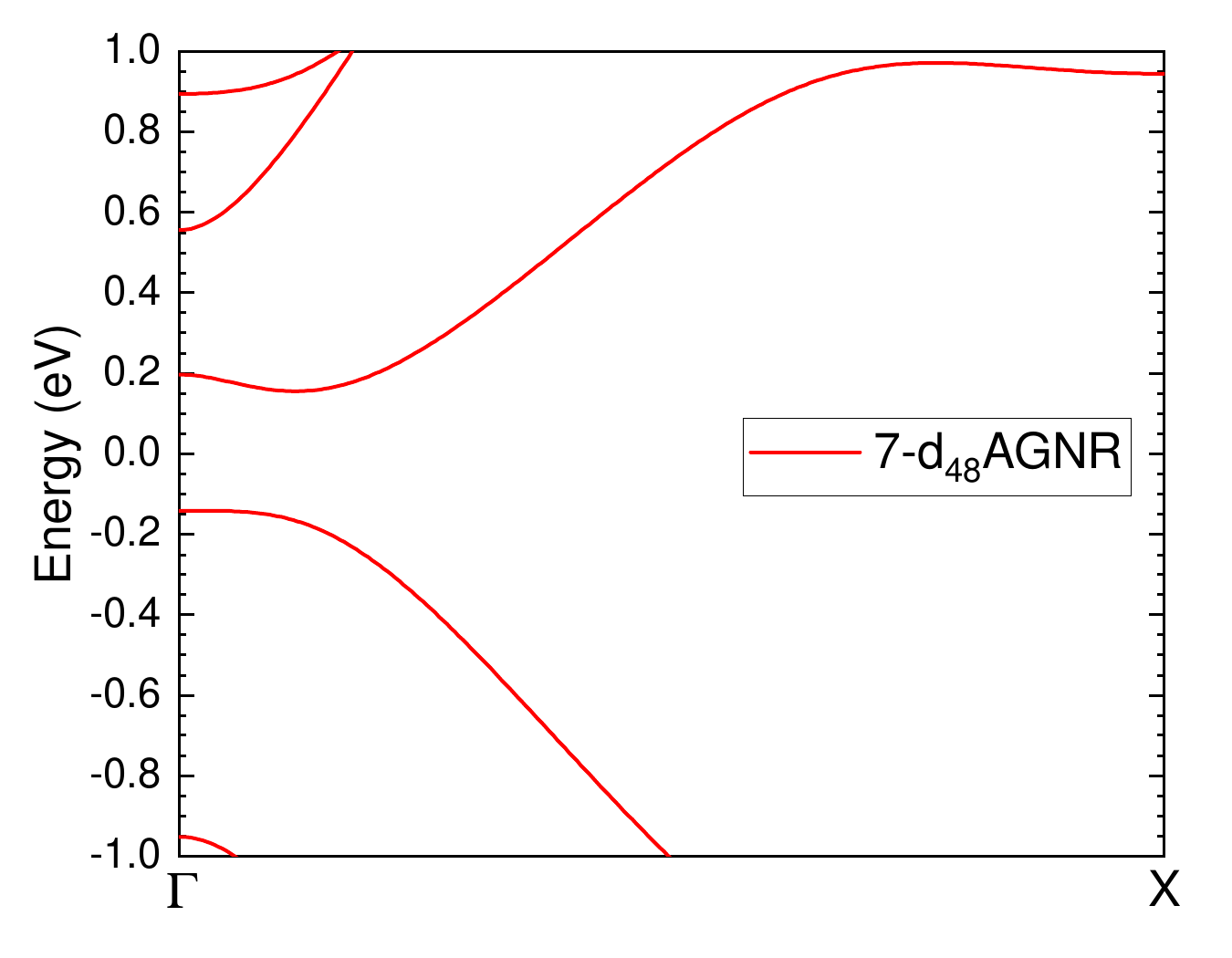}
\includegraphics*[width=0.28\columnwidth]{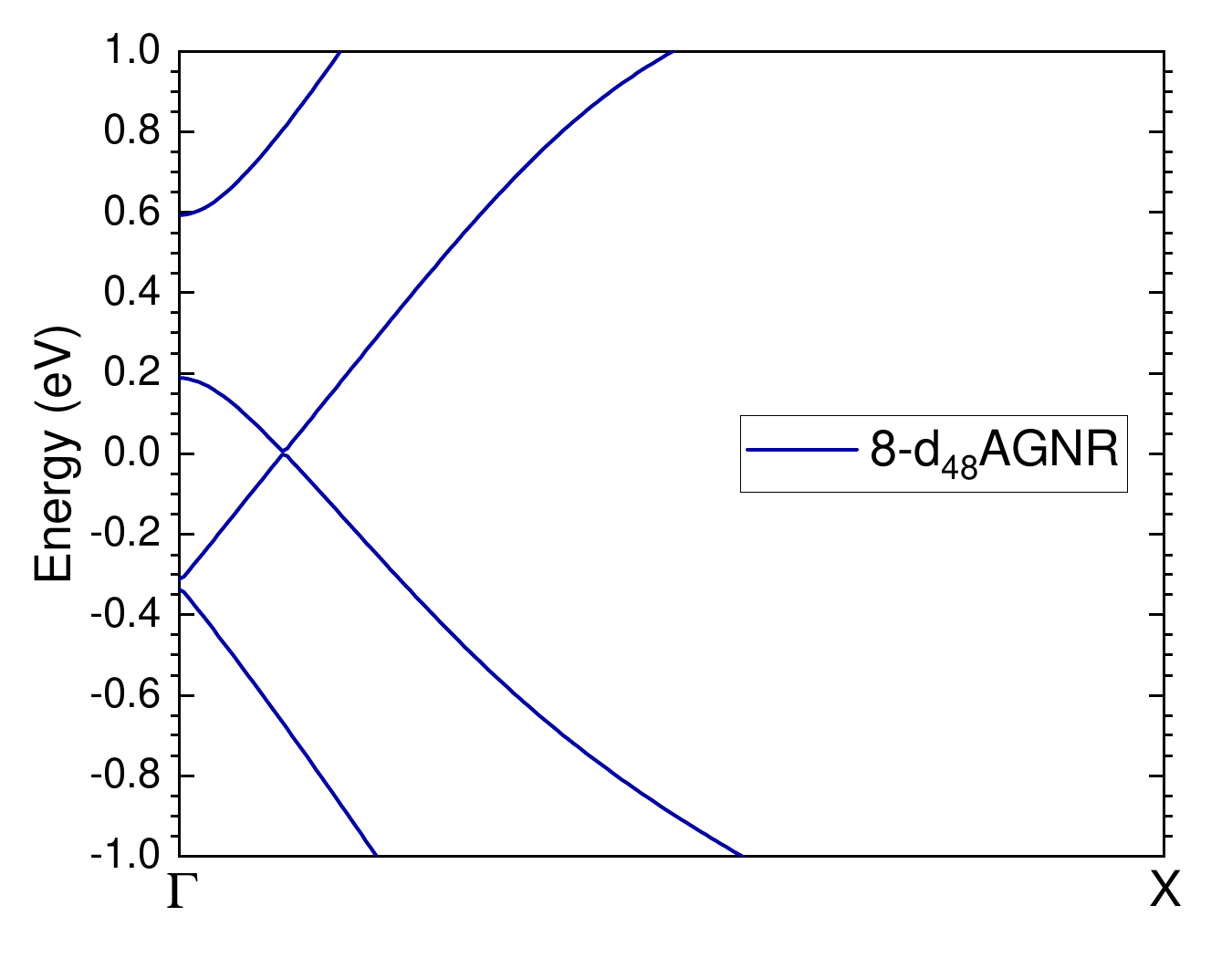}
\includegraphics*[width=0.28\columnwidth]{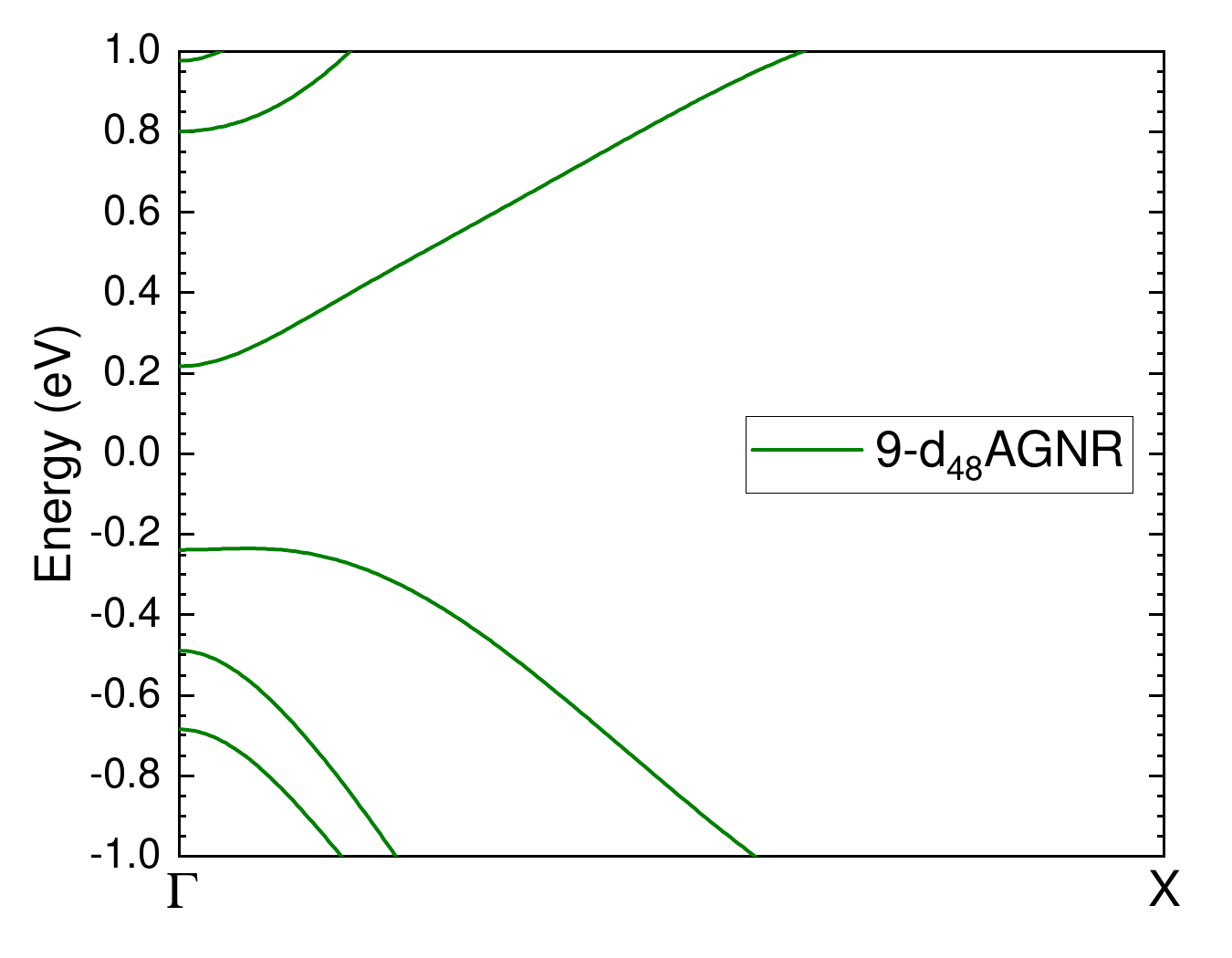}
\includegraphics*[width=0.28\columnwidth]{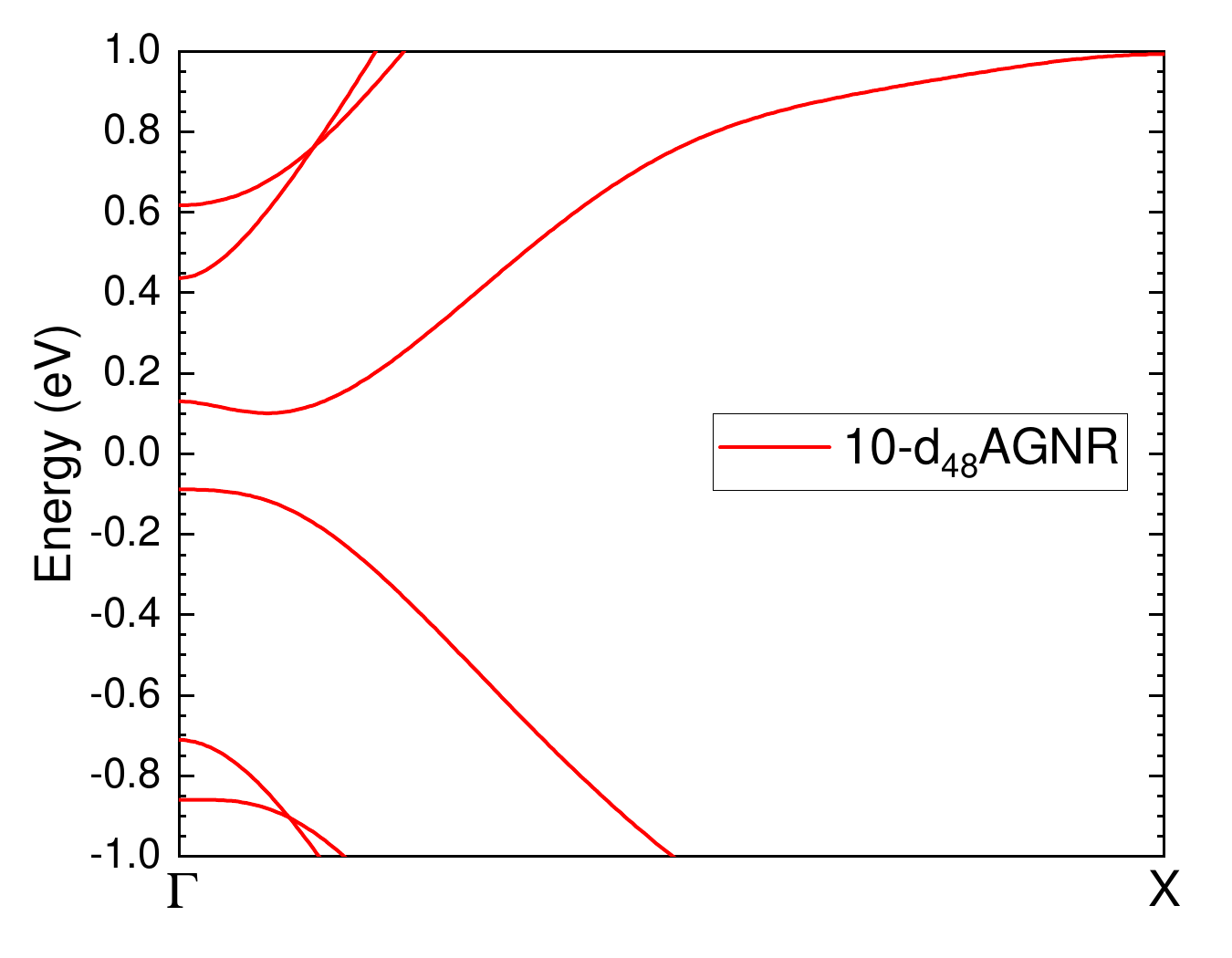}
\includegraphics*[width=0.28\columnwidth]{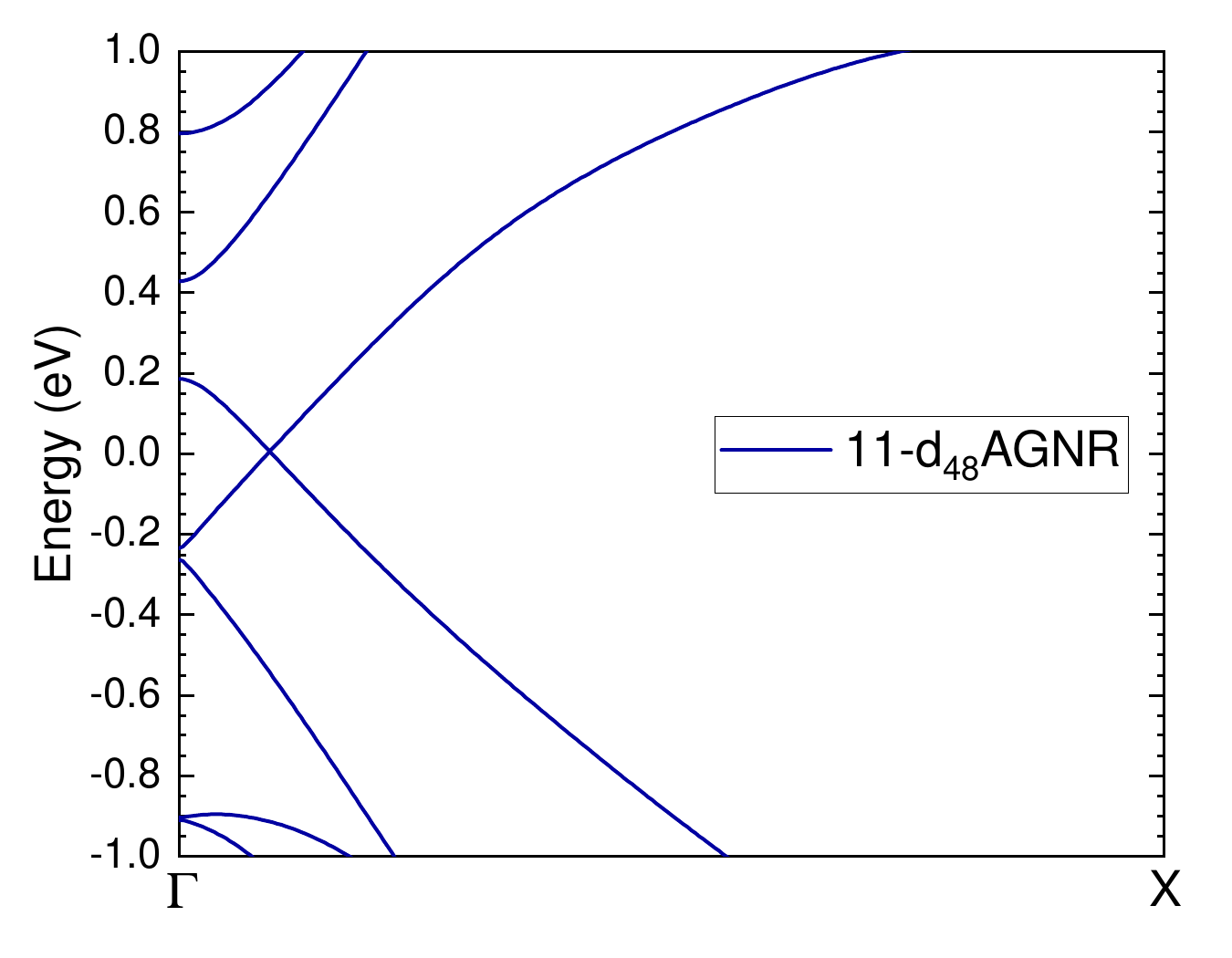}
\includegraphics*[width=0.28\columnwidth]{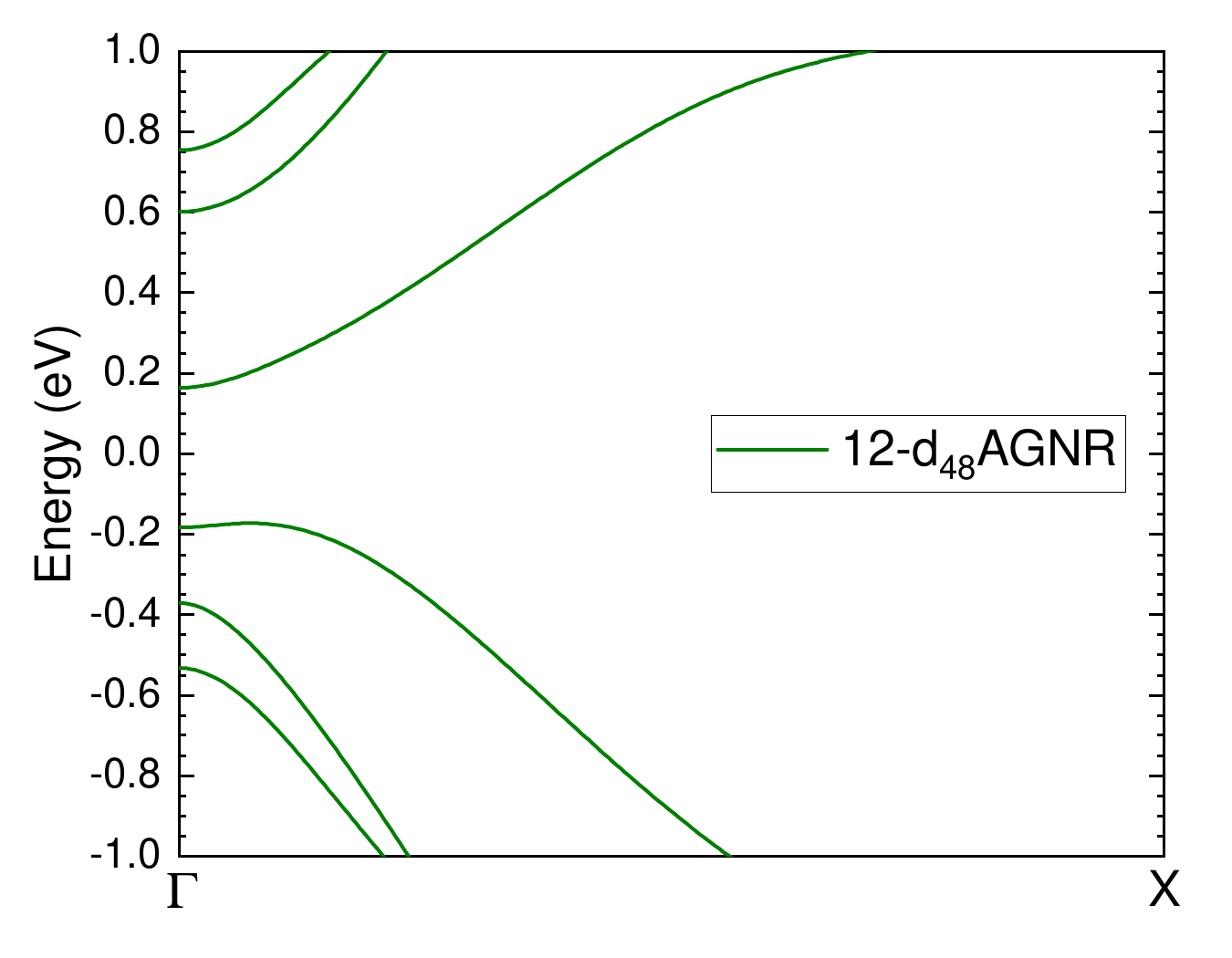}
\includegraphics*[width=0.28\columnwidth]{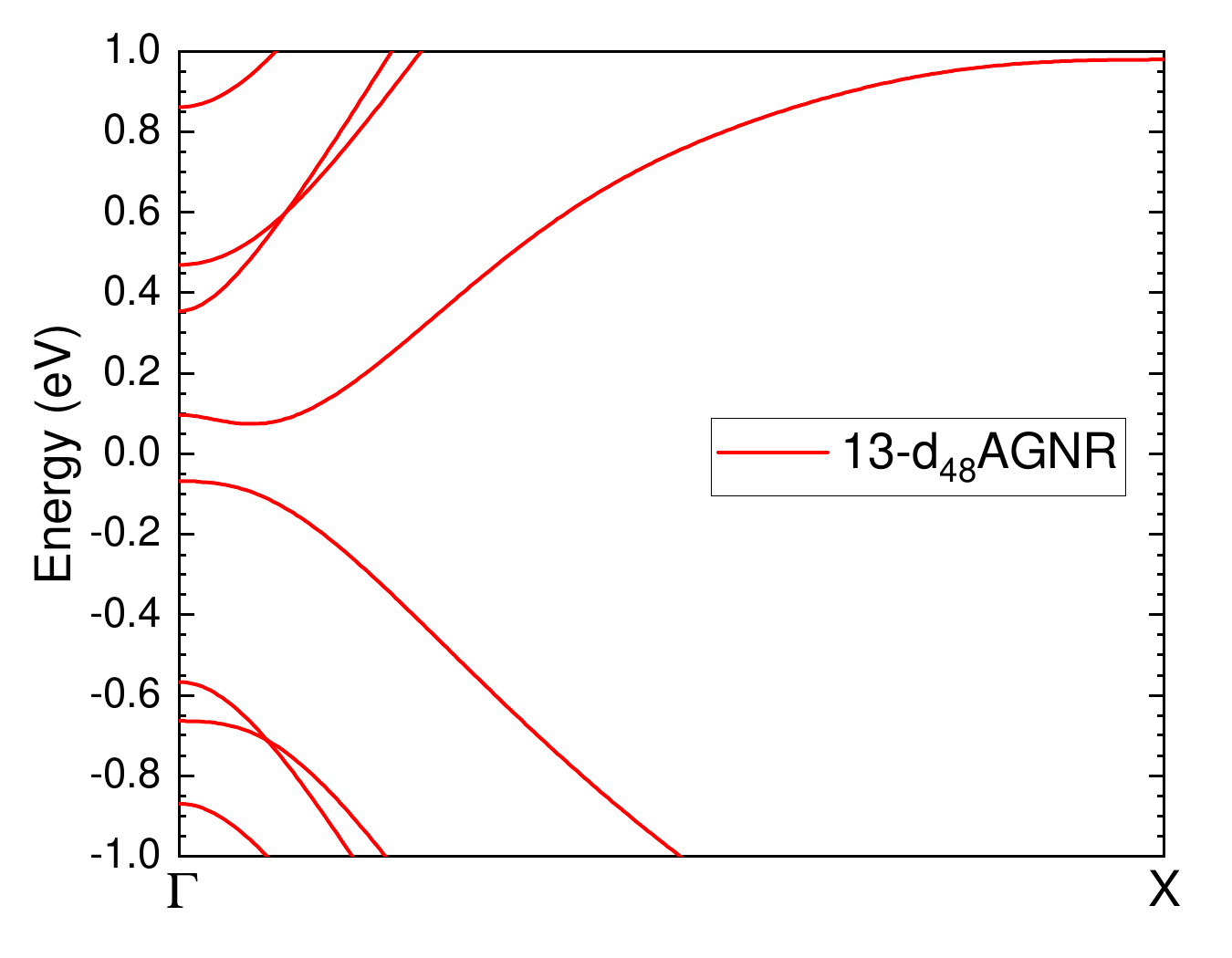}
\includegraphics*[width=0.28\columnwidth]{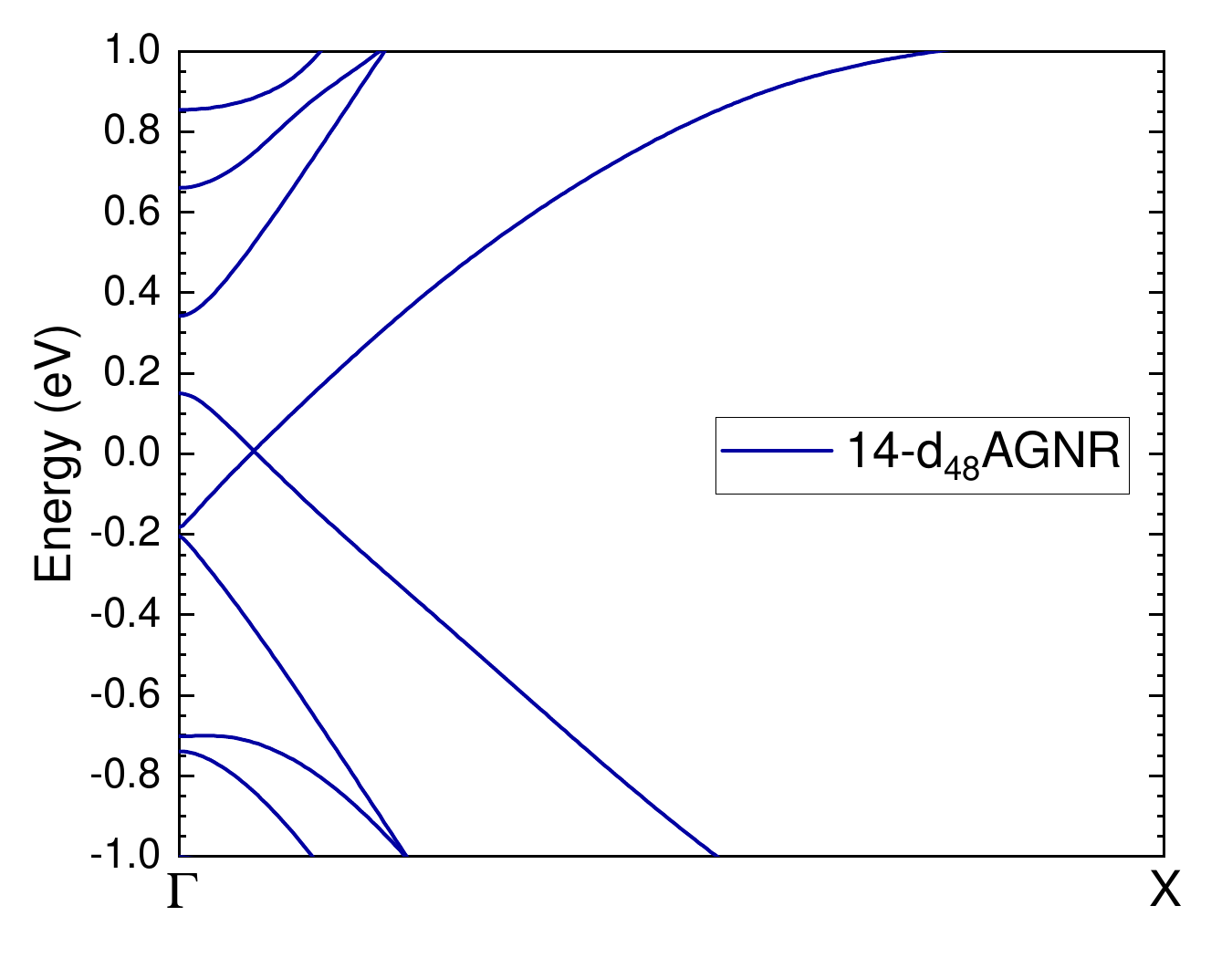}
\includegraphics*[width=0.28\columnwidth]{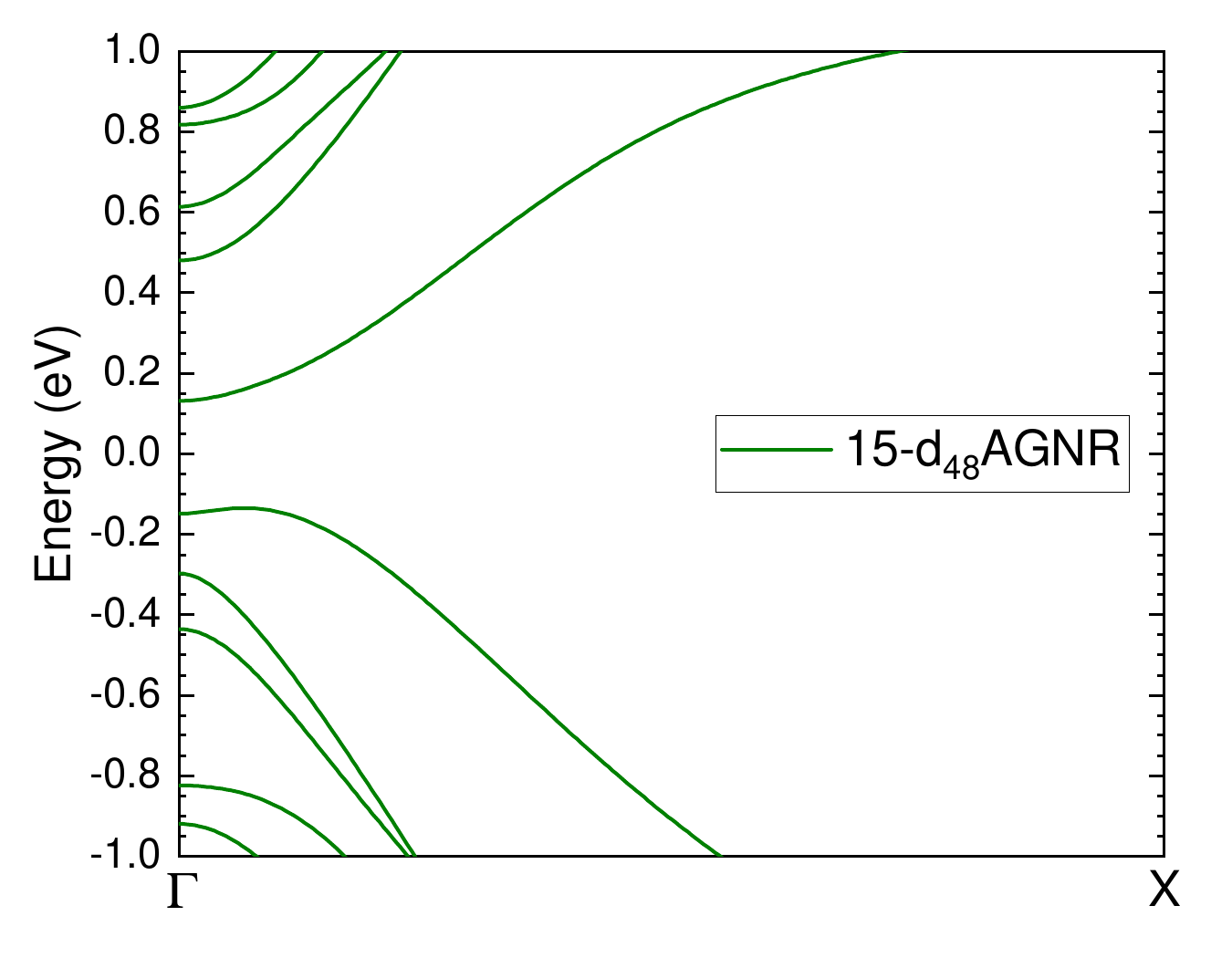}
\includegraphics*[width=0.28\columnwidth]{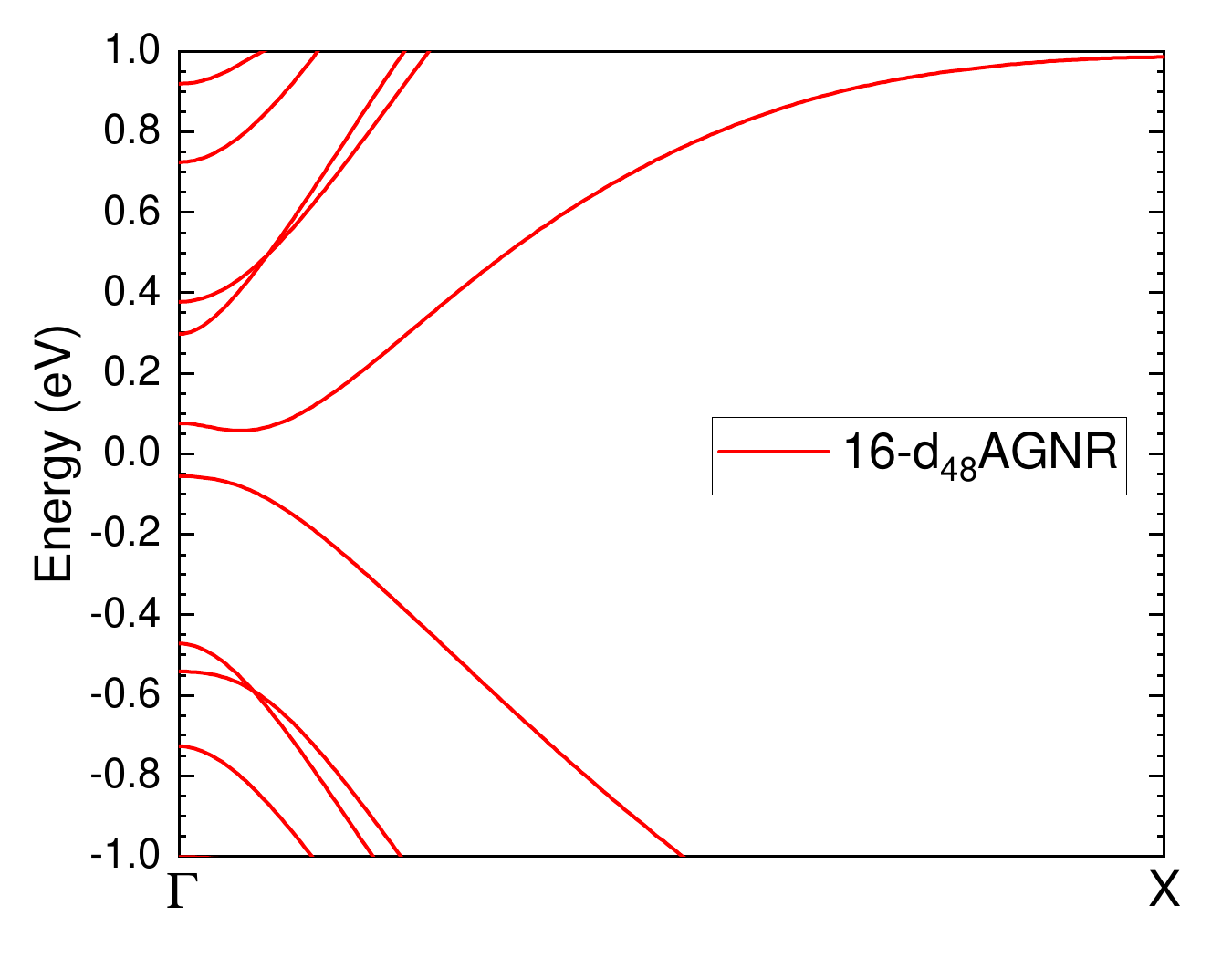}
\includegraphics*[width=0.28\columnwidth]{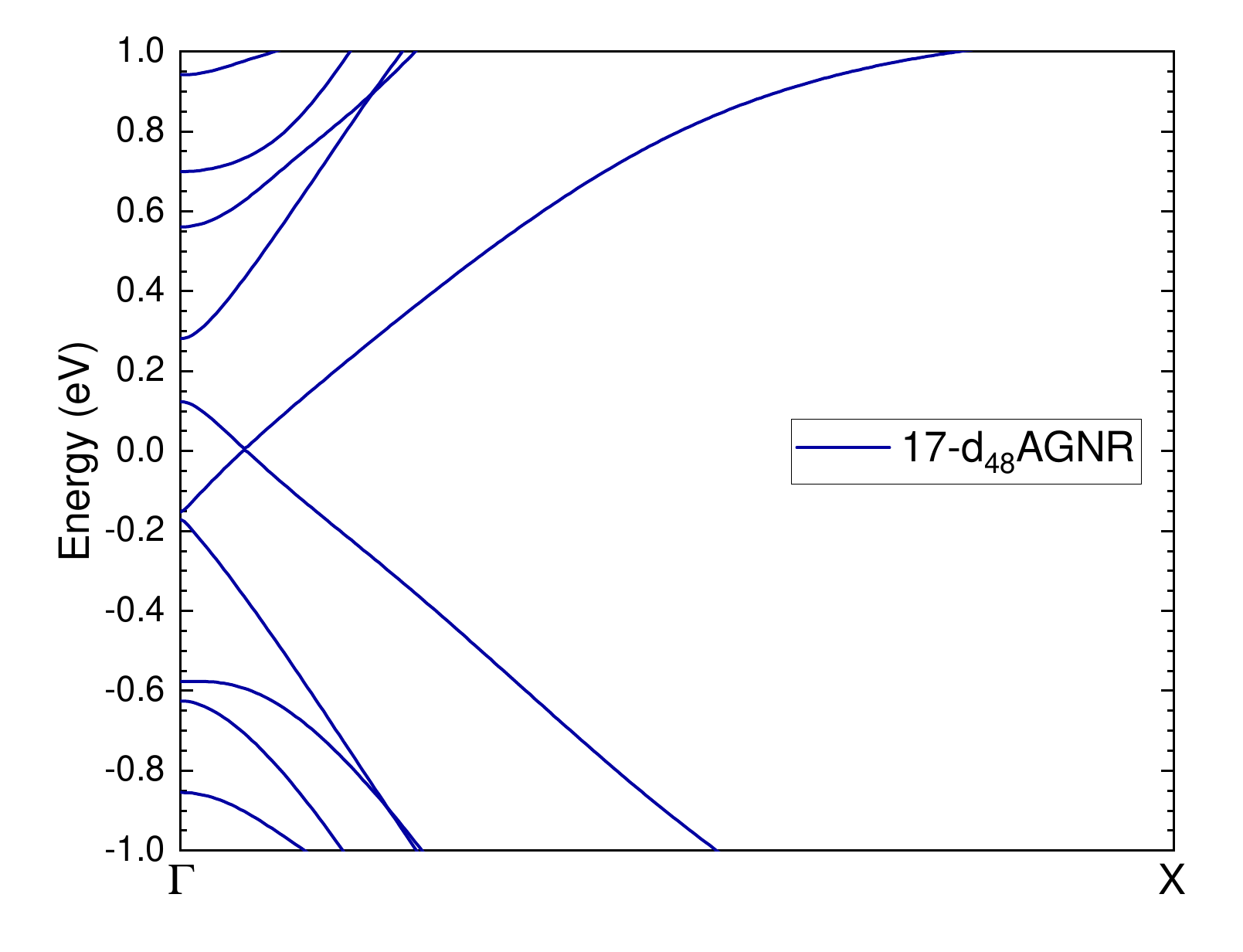}
\caption{\label{fig:bands_3-17} Electronic bandstructures of $N$-d$_{48}$AGNR for $N=3-17$. The left (green color), middle (red color) and right (blue color) columns show the bandstructures of family I, II and III nanoribbons, respectively. In all plots, the zero-of-energy was set to the Fermi energy of the system. }
\end{figure}
\FloatBarrier
\begin{figure}
\centering
\includegraphics*[width=0.28\columnwidth]{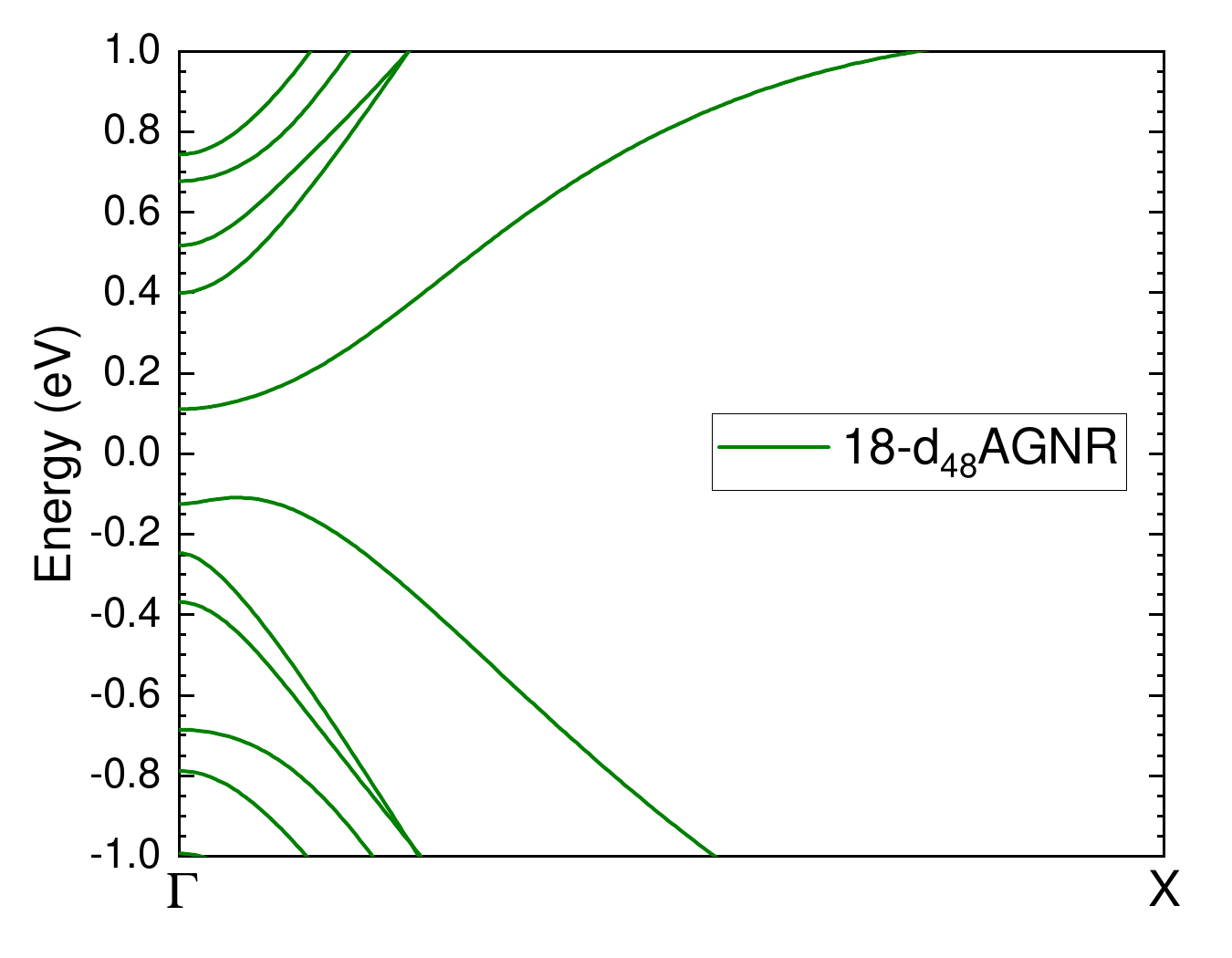}
\includegraphics*[width=0.28\columnwidth]{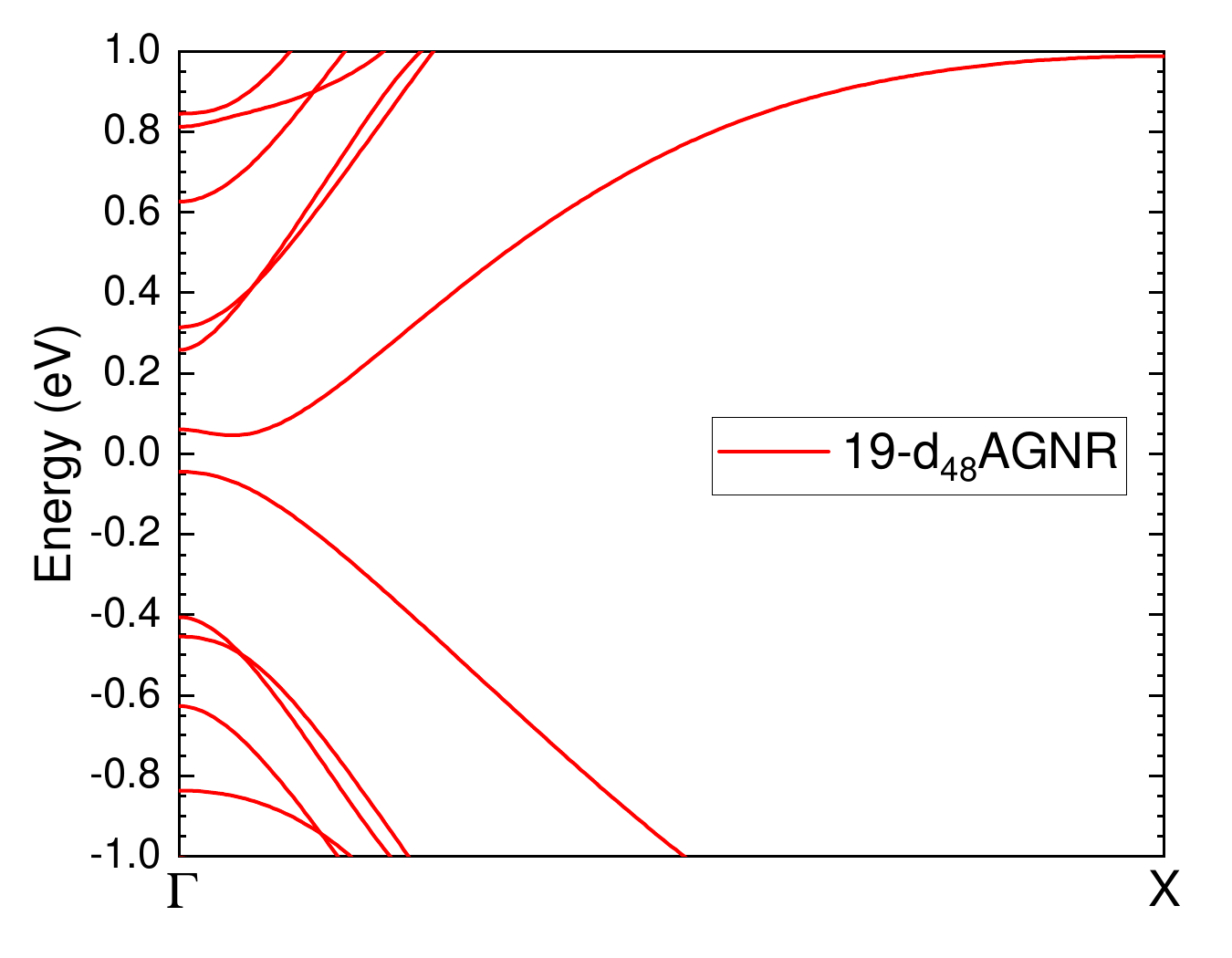}
\includegraphics*[width=0.28\columnwidth]{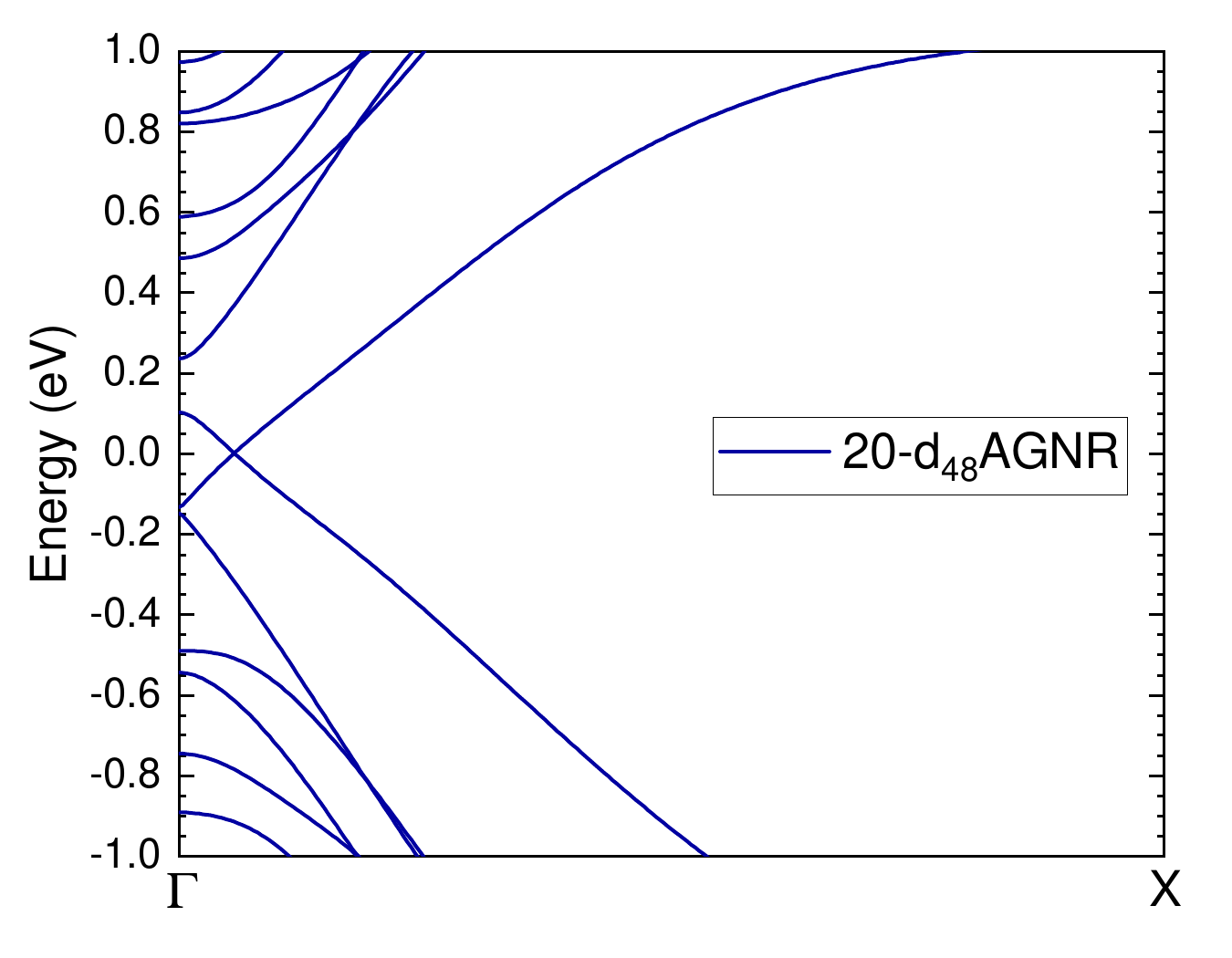}
\includegraphics*[width=0.28\columnwidth]{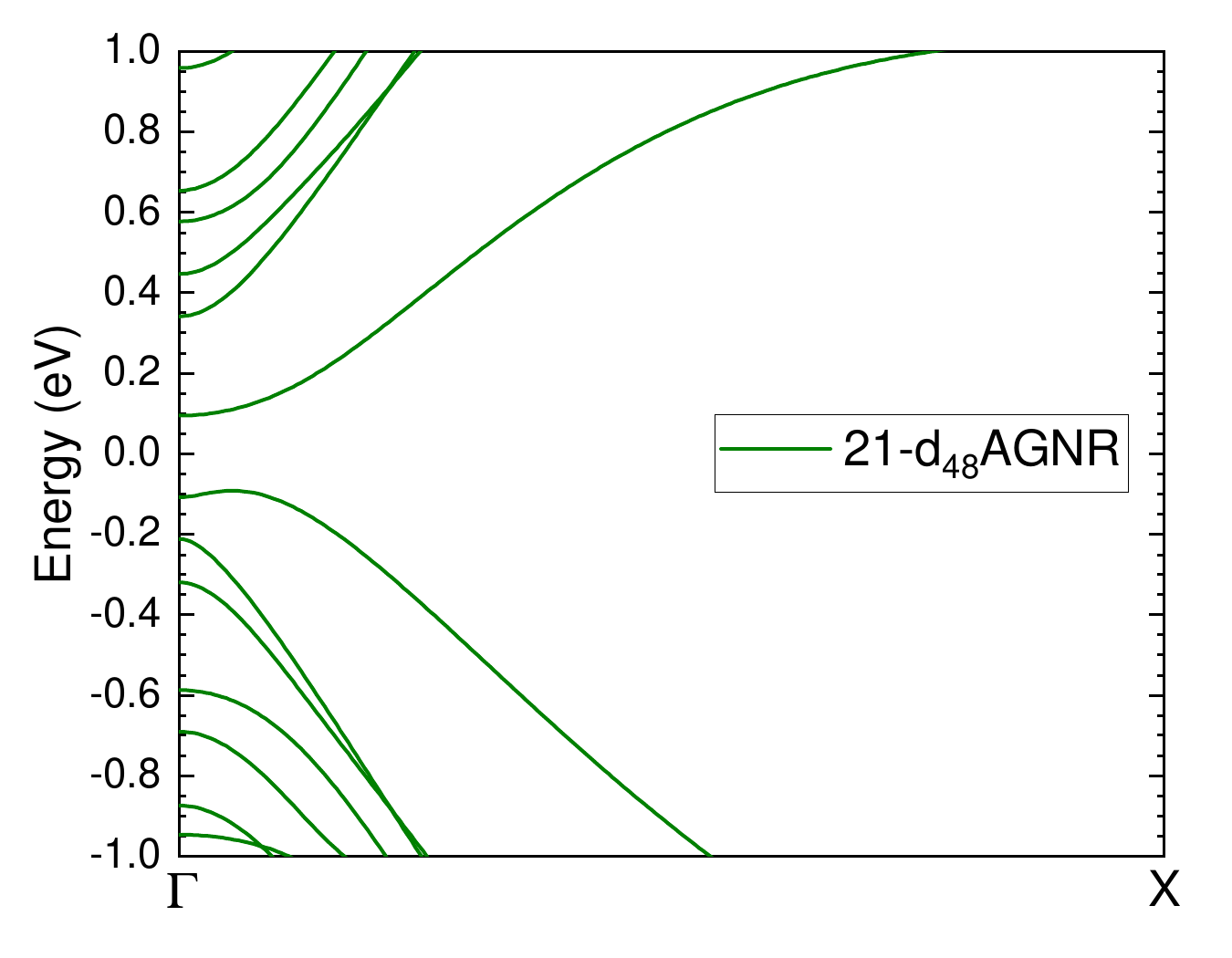}
\includegraphics*[width=0.28\columnwidth]{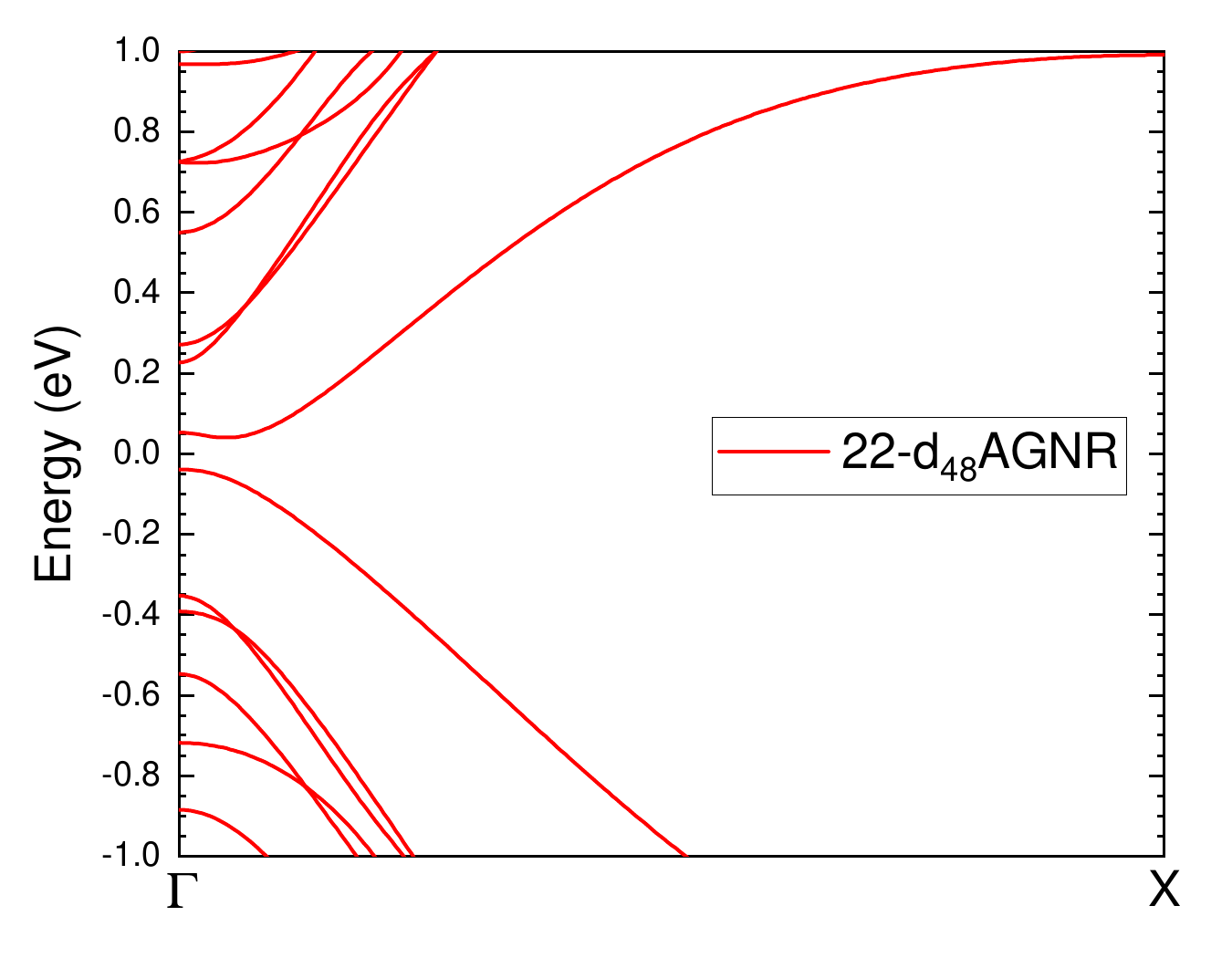}
\includegraphics*[width=0.28\columnwidth]{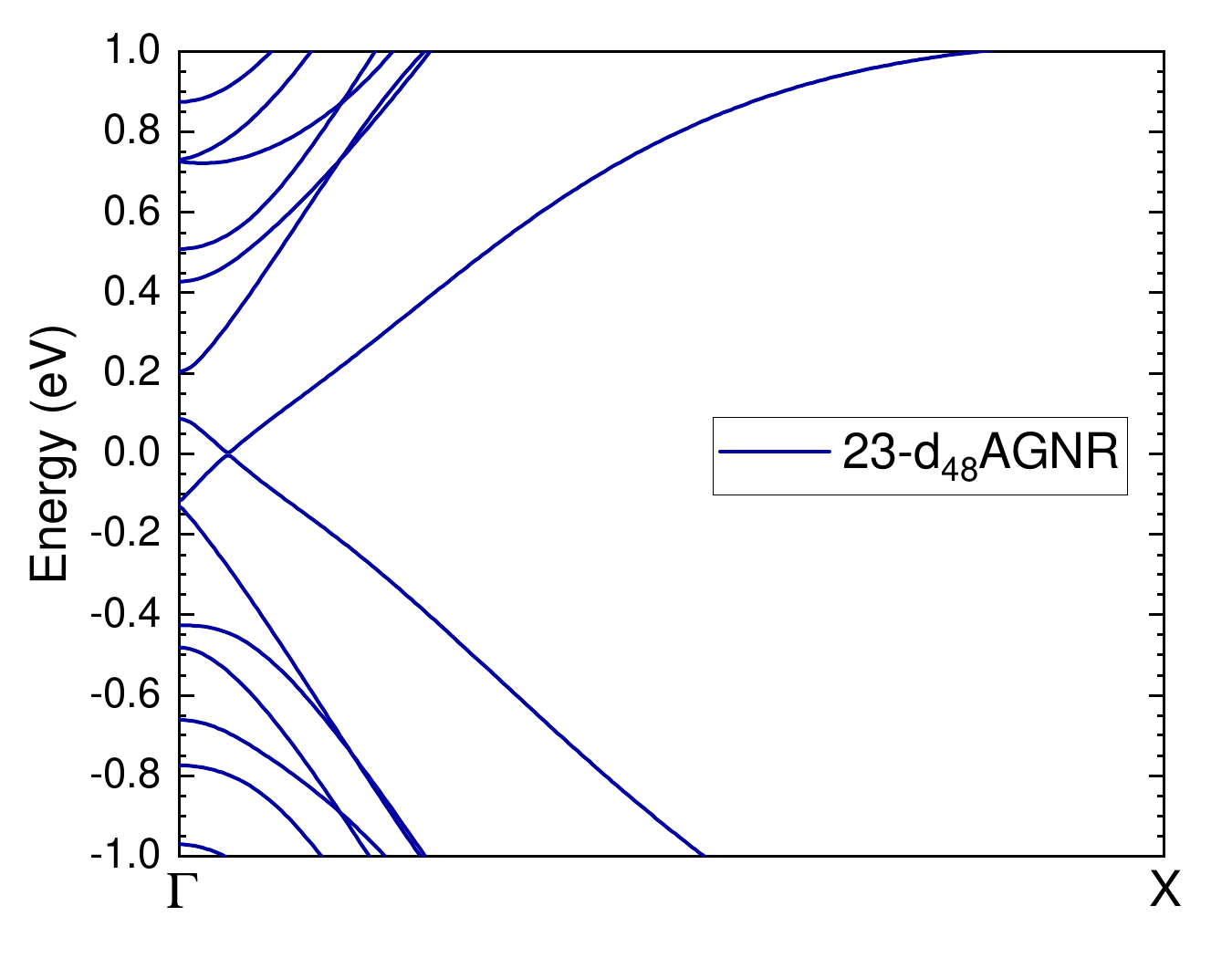}
\includegraphics*[width=0.28\columnwidth]{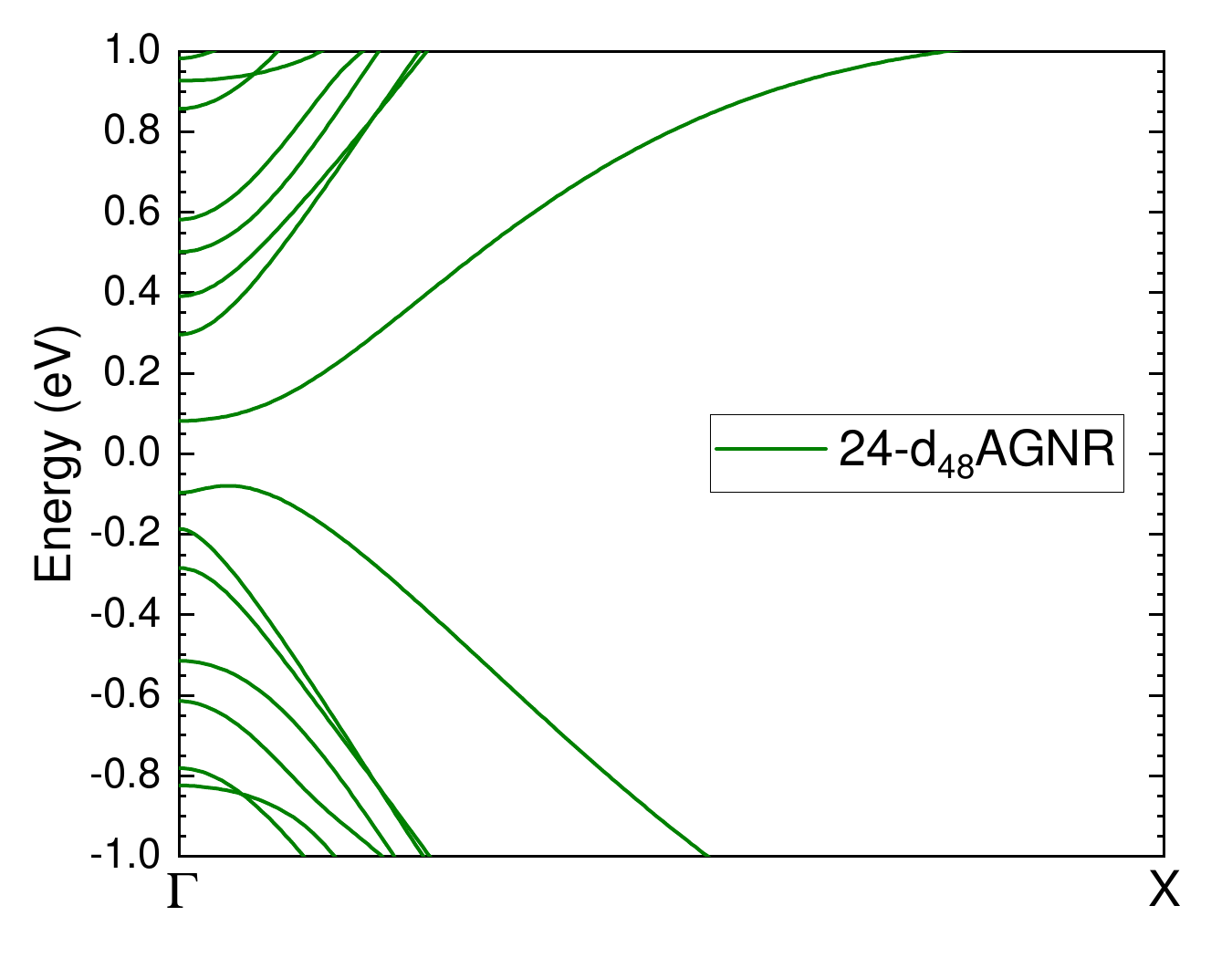}
\includegraphics*[width=0.28\columnwidth]{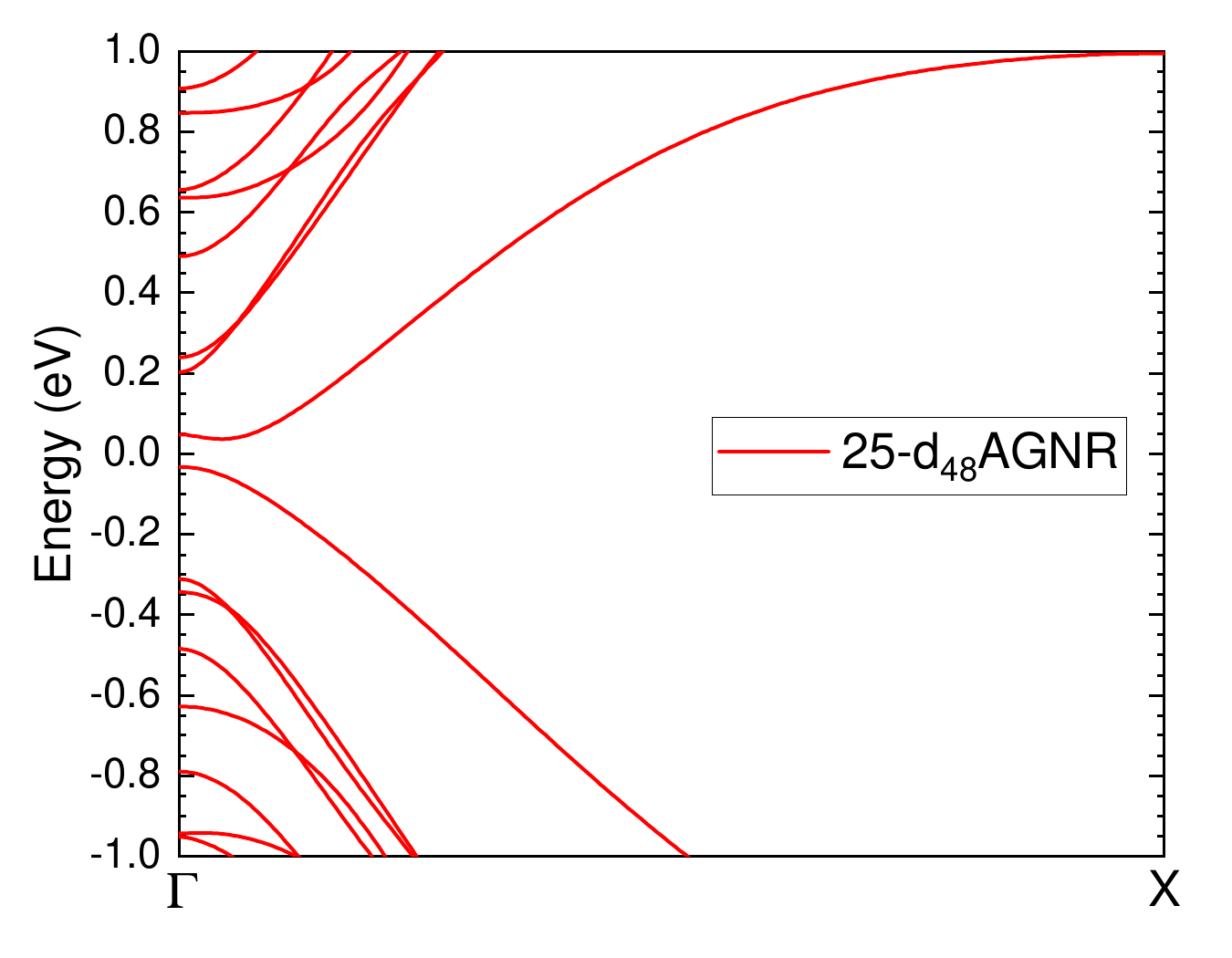}
\includegraphics*[width=0.28\columnwidth]{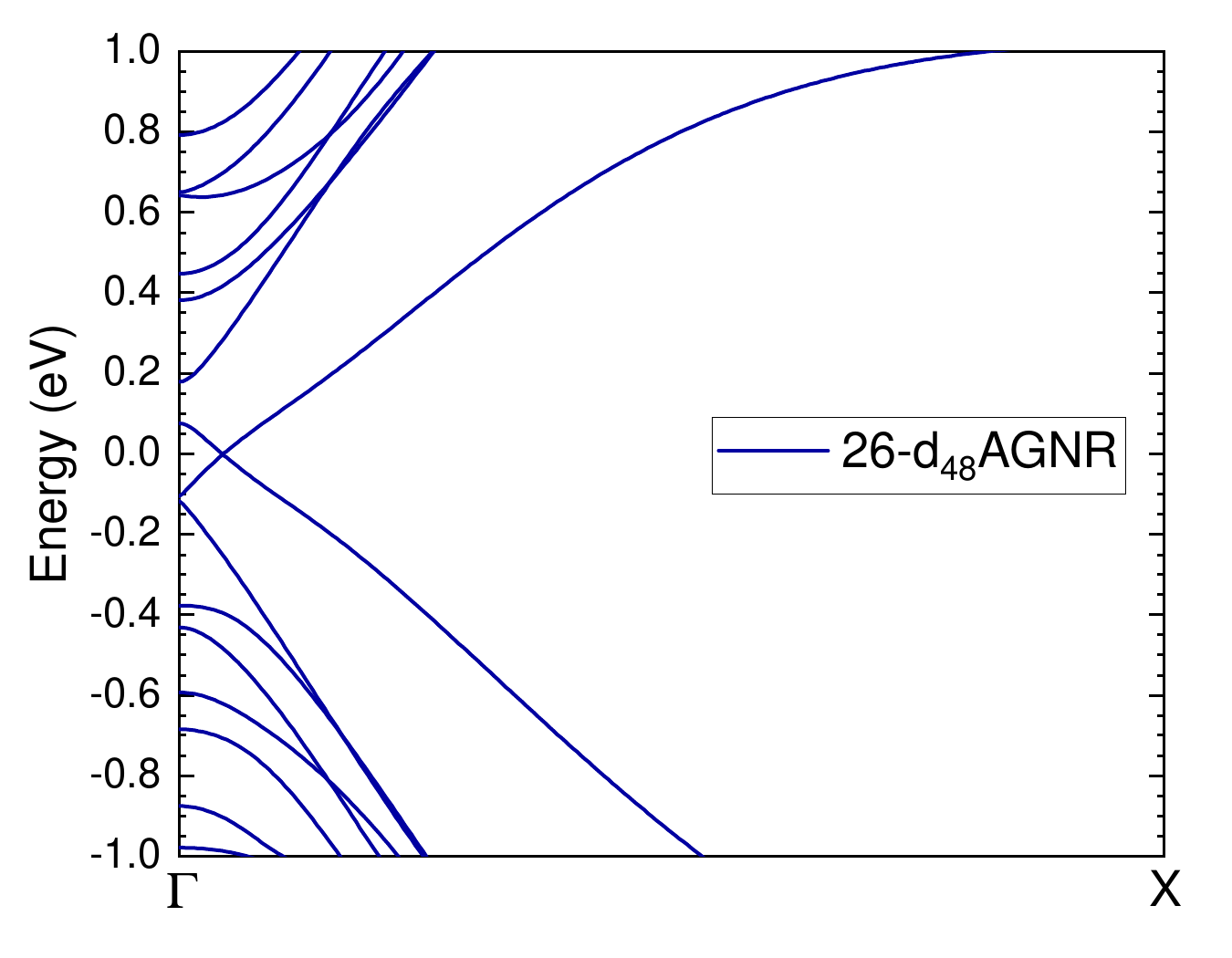}
\includegraphics*[width=0.28\columnwidth]{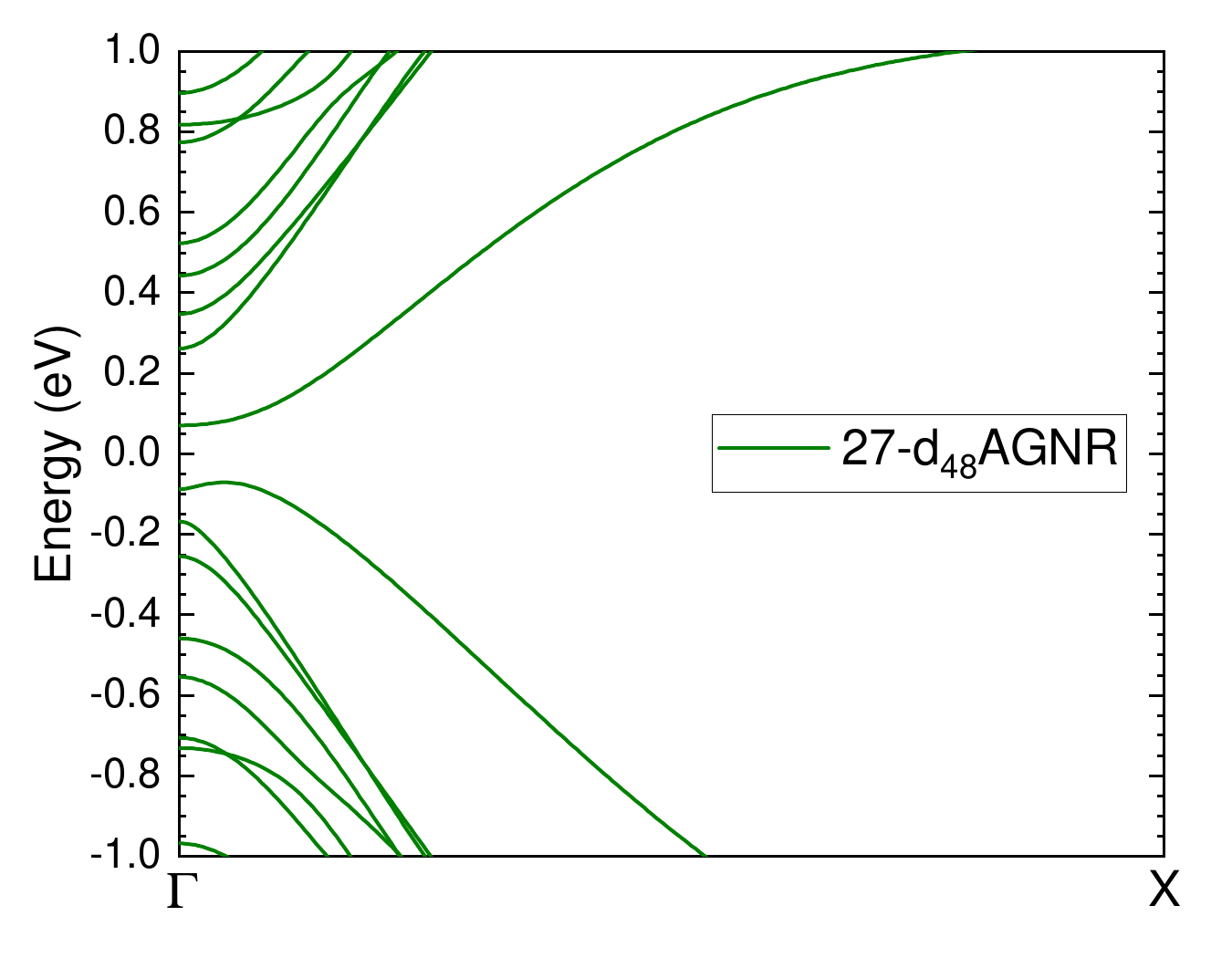}
\includegraphics*[width=0.28\columnwidth]{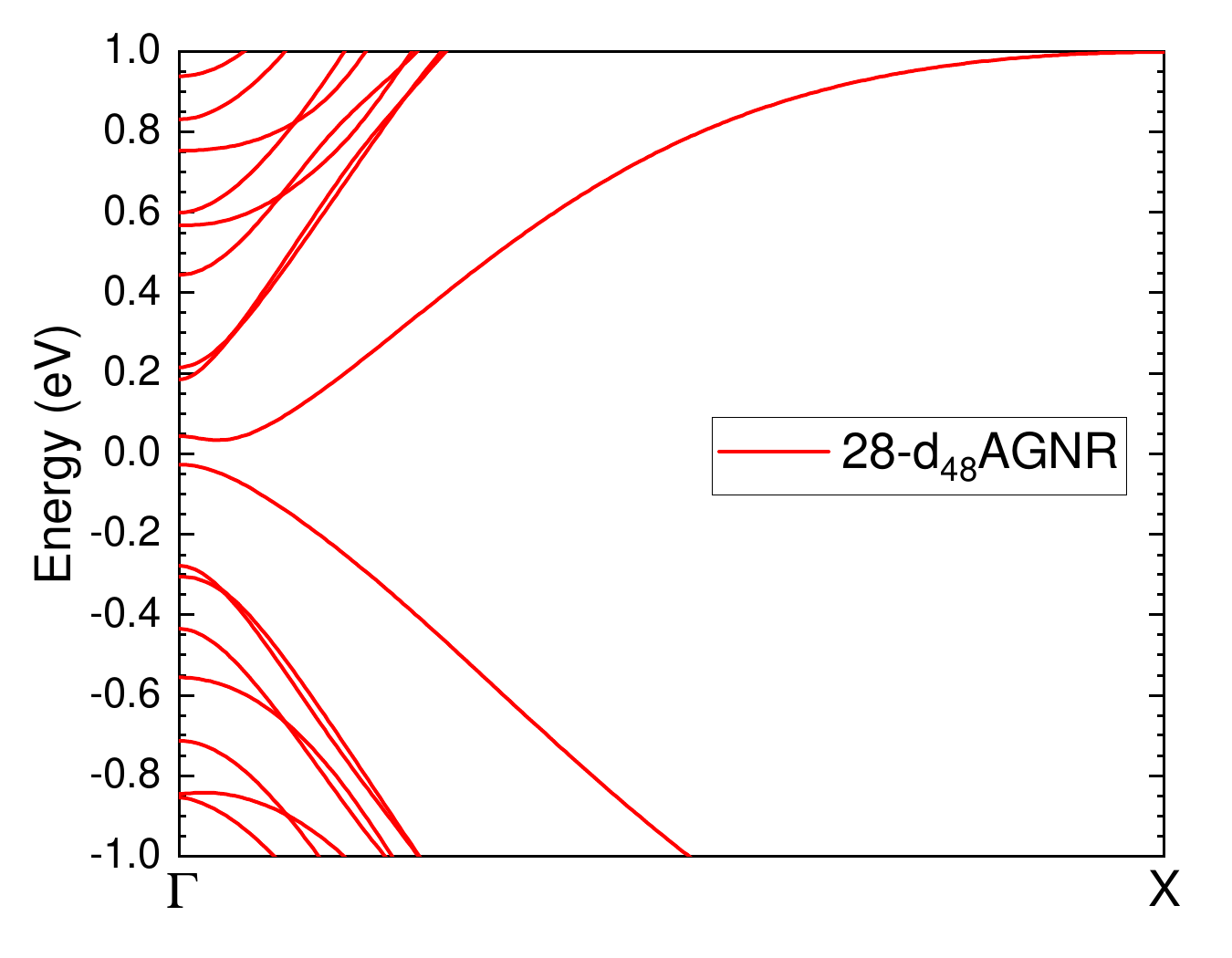}
\includegraphics*[width=0.28\columnwidth]{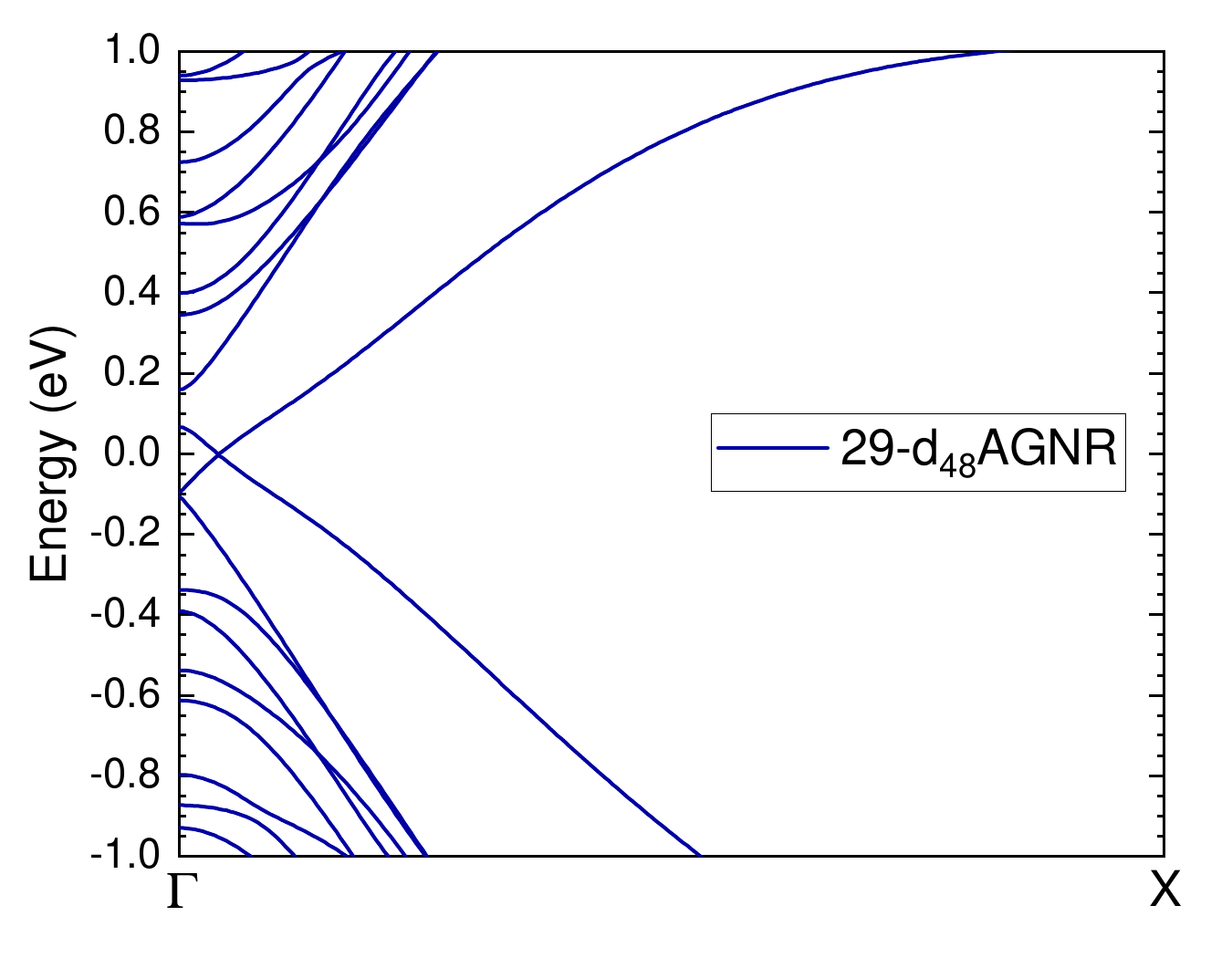}
\includegraphics*[width=0.28\columnwidth]{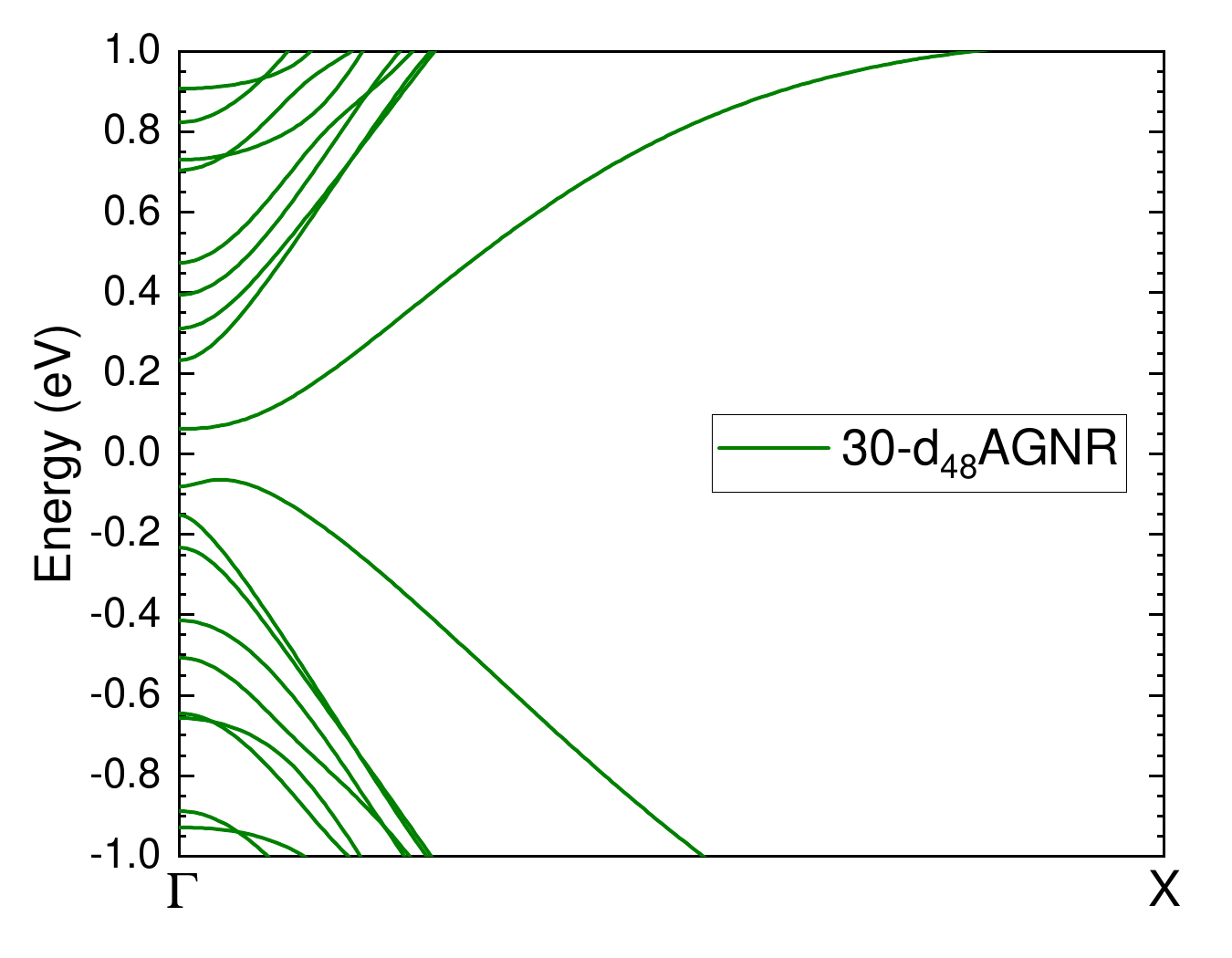}
\includegraphics*[width=0.28\columnwidth]{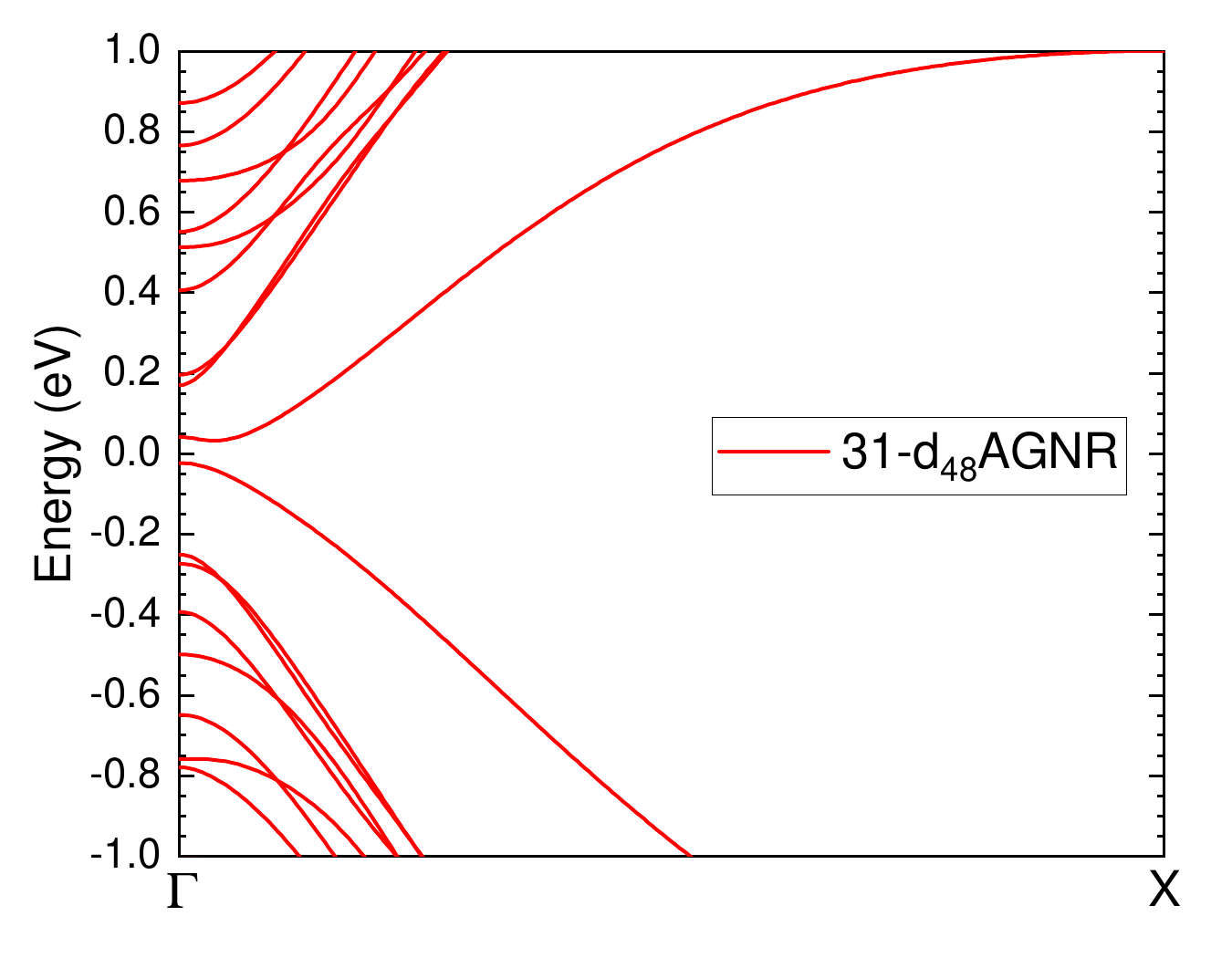}
\includegraphics*[width=0.28\columnwidth]{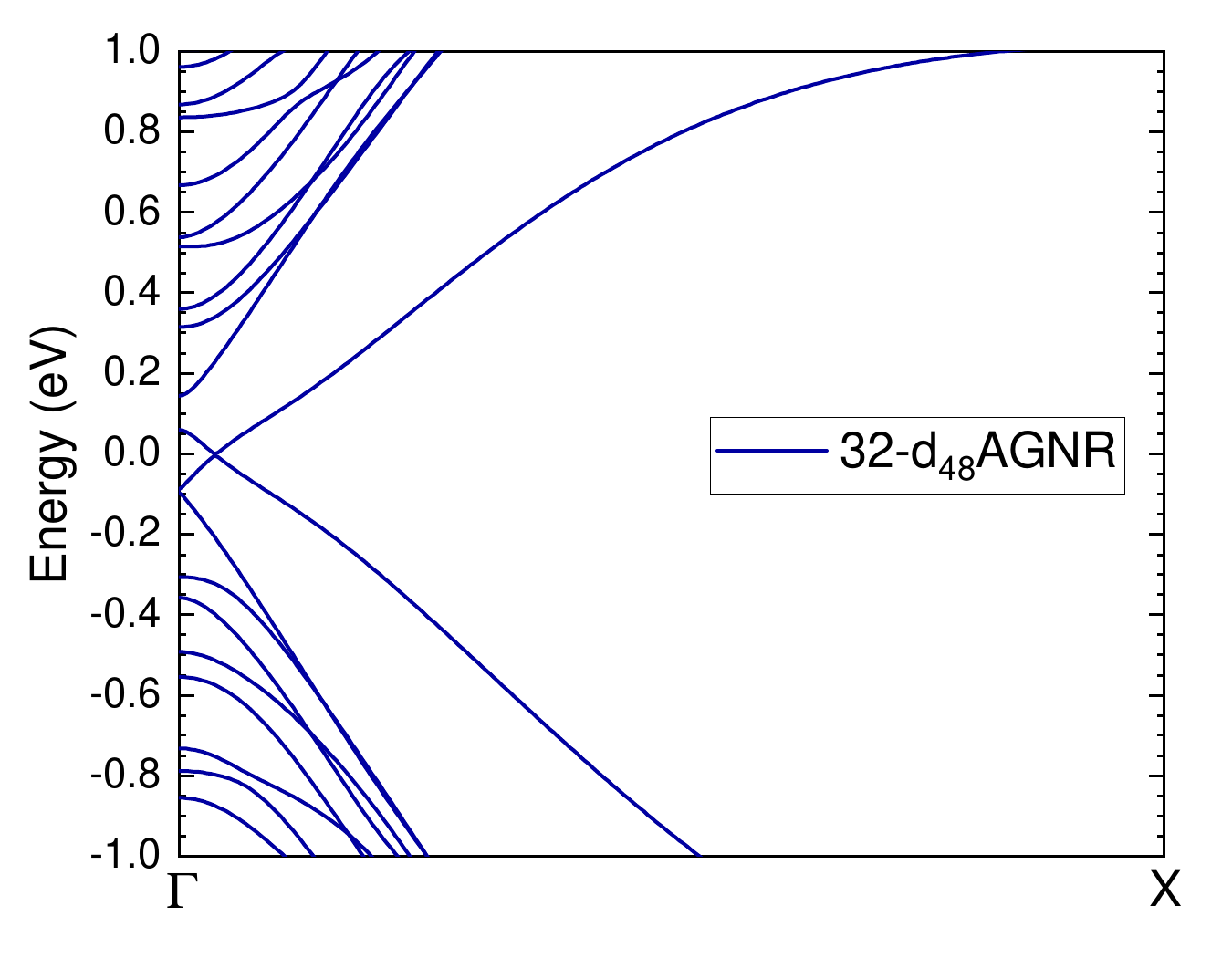}
\caption{\label{fig:bands_18-32} Like Fig.~\ref{fig:bands_3-17}, but for $N=18-32$. }
\end{figure}

\FloatBarrier
\newpage

\section{Pristine AGNR wavefunctions}
\begin{figure}
\centering
\includegraphics*[width=0.45\columnwidth]{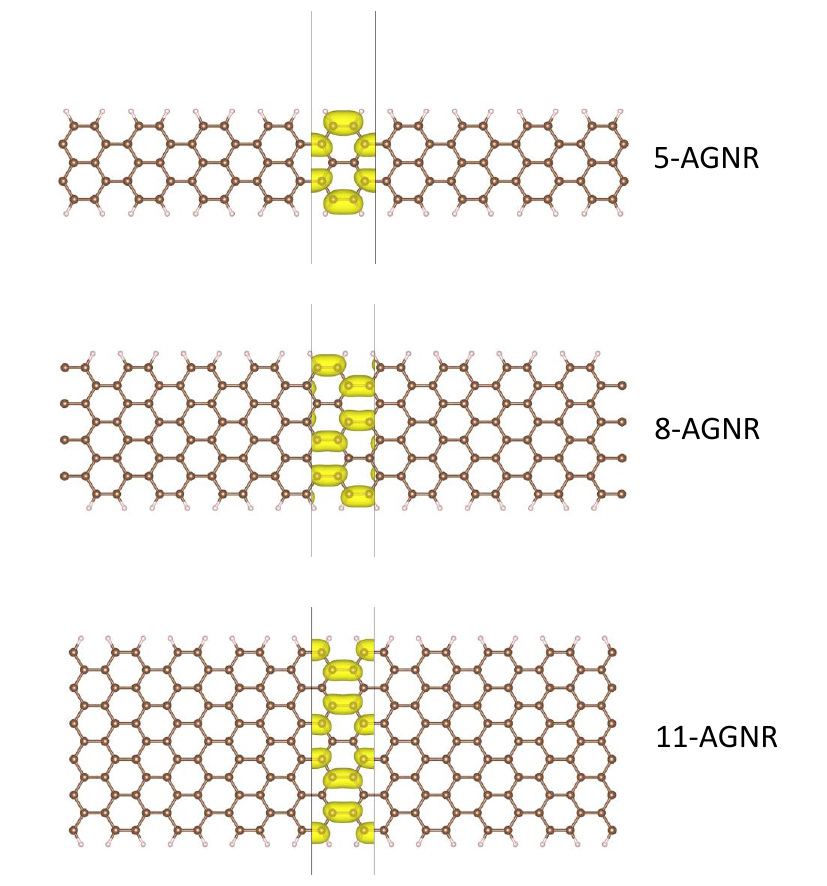}
\caption{\label{fig:nodal_lines} Wavefunctions at valence band top at $\Gamma$ point for $3p+2$ family AGNRs of increasing width.}
\end{figure}

Figure~\ref{fig:nodal_lines} shows the wavefunctions of the valence band maximum at the $\Gamma$ point of various AGNRs from the $3p+2$ family. The wavefunctions consist of armchair-shape lines of probability density along the nanoribbon axis, which are broken by 'nodal lines' with vanishing probability density.


\end{document}